\documentclass[a4paper,twocolumn]{esapub2005} 
\usepackage{times,graphicx}
\usepackage{natbib}

\title{A line profile analysis of the pulsating 
red giant star $\epsilon$~Ophiuchi~(G9.5III)}

\author[1,2]{S. Hekker}
\author[2,3]{C. Aerts}
\author[2]{J. De Ridder}
\author[2,4]{F. Carrier}
\affil[1]{Leiden Observatory, P.O. Box 9513, 2333 RA Leiden, The Netherlands, E-mail: saskia@strw.leidenuniv.nl }
\affil[2]{Instituut voor Sterrenkunde, Katholieke Universiteit Leuven, Celestijnenlaan 200 B, B-3001 Leuven, Belgium}
\affil[3]{Department of astrophysics, University of Nijmegen, P.O. Box 9010, 6500 GL Nijmegen, The Netherlands}
\affil[4]{Observatoire de Gen\`eve, 51 Chemin de Maillettes, 1290 Sauverny, Switzerland}

\begin{document}


\maketitle

\begin{abstract}
So far, solar-like oscillations have been studied using radial velocity and/or
light curve variations, which reveal frequencies of the oscillation modes.
Line-profile variations, however, are also a valuable diagnostic to characterise
radial and non-radial oscillations, including frequencies, amplitudes, the
spherical mode wavenumbers $(\ell,m)$ and the stellar inclination angle.  Here
we present a line profile analysis of $\epsilon$ Ophiuchi, which is a pulsating
red giant.  The main differences compared to previous line profile analyses done
for heat-driven oscillations are the small amplitudes and the predicted short
damping and re-excitation times in red giants.

Two line diagnostics have been tested to see whether these are sensitive to the
small line profile variations present in red giants. In addition, line profiles
have been simulated with short damping and re-excitation times and are compared
with the observations.  This comparison reveals that non-radial modes are
detected in the observed line profile variations of $\epsilon$ Ophiuchi. This is
rather surprising, as theoretical predictions favours the occurrence of radial
modes.
\end{abstract}

\section{Introduction}

The discovery of oscillations in the sun, about five decades ago, was the onset
of seismology of stars. Oscillations probe the interior of the stars and are
therefore a direct means to determine their internal structure. Solar-like
oscillations are excited by turbulent convection near the surface of cool stars
of spectral type F, G, K or M. They show radial velocity variations with
amplitudes of typically a few cm\,s$^{-1}$ to a few m\,s$^{-1}$, and with
periods ranging from a few minutes for main sequence stars to about half an hour
for subgiants and a couple of hours for giants.

During the last decade, the refinement of the techniques to perform very
accurate radial velocity observations, made it also possible to observe
solar-like oscillations in distant stars.  Recently, these solar-like
oscillations are indeed observed in several red giants, for example in $\xi$
Hydrae \cite{frandsen2002}, $\epsilon$ Ophiuchi \cite{deridder2006b} and $\eta$
Serpentis \cite{carrier2006}.

This proceedingspaper describes observations as well as a frequency analysis of
different parameters and a line profile analysis of $\epsilon$ Ophiuchi. It is
organised as follows. In Section 2, we start with a discussion on the target
selection, followed in Section 3 by the description of the observations. In
Section 4 the data analysis is described, i.e. first, the frequency
determination of different oscillation diagnostics and subsequently the mode
identification using these frequencies.  In Section 5 we show simulations,
including the short damping and re-excitation times present in solar-like
oscillations of red giants \cite{stello2004}.  Finally, in Section 6 we draw
some conclusions.

\section{Target selection}

\begin{table}
\caption{Basic stellar parameters of $\epsilon$ Ophiuchi taken from
\cite{deridder2006b}:
Effective temperature ($T_{\mathrm{eff}}$) in Kelvin, rotational velocity
($\upsilon \mathrm{sin}i$) in km\,s$^{-1}$, parallax ($\pi$) in mas, distance
($d$) in pc, the apparent magnitude ($m_{v}$) and absolute magnitude ($M_{V}$)
in the V band.}
\label{propstar}
\centering
\begin{tabular}{lc}
\hline\hline
parameter & $\epsilon$ Ophiuchi\\
\hline
$T_{\mathrm{eff}}$ [K] & $4887 \pm 100$ \\
$\upsilon \mathrm{sin}i$ [km\,s$^{-1}$] & $3.4 \pm 0.5$\\
$\pi$ [mas] & $30.34\pm0.79$ \\
$d$ [pc] & $33.0\pm 0.9$ \\
$m_{v}$ [mag] & $3.24\pm0.02$ \\
$M_{V}$ [mag] & $0.65\pm0.06$\\
\hline
\end{tabular}
\end{table}

In order to obtain radial velocity variations with an accuracy of order
m\,s$^{-1}$, the target selection as well as the observing strategy are
important. First, the stars have to be bright enough to obtain high signal to
noise ratios to detect the signal. Secondly, the rotational velocities have to
be low in order to have narrow spectral lines, which are needed for precise
radial velocity measurement. Also other infering phenomena like star spots or
companions should be avoided.

For the observation strategy we have to take into account that the expected (and
observed) period of the solar-like oscillations in red giants is of order a
couple of hours. This means that the observation times should be of order
minutes, in order not to average over a large part of the oscillation
cycle. Furthermore single site observations always introduce one day aliases due
to the diurnal cycle. Multi site observations, or continuous observations from
space can significantly reduce this aliasing effect.

The G9.5III giant $\epsilon$ Ophiuchi was selected for observations of
solar-like oscillations, because of its brightness, $m_{v}$ = 3.24 mag, its low
rotational velocity, $v \sin i$ = 3.4 km~s$^{-1}$, and the low declination,
$\delta$ = -04 41 33.0 \cite{barban2004}. The latter was very important to be
able to observe the star from both La Silla, Chile and Observatoire de Haute
Provence, France.  More basic stellar parameters of $\epsilon$ Ophiuchi are
shown in Table~\ref{propstar}. Furthermore the Hipparcos catalog \cite{esa1997}
gives a photometric variation of less than 0.06 mag and the star is mentioned in
the '1988 revised MK spectral standards for stars G0 and later'
\cite{keenan1988}. Therefore the presence of starspots or companions is not very
likely.

\section{Observations}

The observations of $\epsilon$ Ophiuchi were performed in a bi-site campaign
with the fibre-fed \'echelle spectrograph CORALIE, mounted on the Swiss 1.2~m
Euler telescope at La Silla (ESO, Chile), and the fibre-fed \'echelle
spectrograph ELODIE, mounted on the French 1.93~m telescope at the Observatoire
de Haute Provence (France), during the summer of 2003. The spectra range from
387.5~nm to 682~nm in wavelength.  The observation times were kept short enough,
in order not to average over a large part of the oscillation, but long enough to
reach a signal to noise ratio of at least 100 at 550~nm.  

The radial  velocities of $\epsilon$  Ophiuchi were determined with  the optimum
weight  method  \cite{bouchy2001a} and  are  available in  \cite{deridder2006b}.
Variations  of order  20~m~s$^{-1}$  due to  stellar  oscillations were  clearly
detected by  the authors. We refer  the reader to  \cite{deridder2006b} for more
information on the radial velocity analysis of the star.

In this work, we use the same data as in \cite{deridder2006b}, but we focus on
the line profile variations of the star rather than on the radial velocity. Such
variations can be derived with high accuracy from a single unblended line
provided that it is not subjected to Stark broadening (see, e.g.,
\cite{aerts2003} and references therein) and that the spectra have a high
signal-to-noise ratio. The latter condition is not fulfilled for our data of
$\epsilon$ Ophiuchi.  Therefore, we perform a line profile analysis from a cross
correlation profile computed of each spectrum. This was done by using a
box-shaped mask \cite{baranne1996}. This cross correlation profile has an
appreciably higher signal-to-noise ratio compared to a single spectral line.
Moreover, \cite{chadid2001} have shown that a line profile analysis in terms of
stellar oscillations is possible using a cross correlation profile rather than
individual lines. This is a similar conclusion to the one by \cite{dall2006},
who showed that the cross-correlation bisector contains the same information as
single-line bisectors.

\section{Data analysis}
\begin{figure}
\centering
\resizebox{\hsize}{!}{\includegraphics{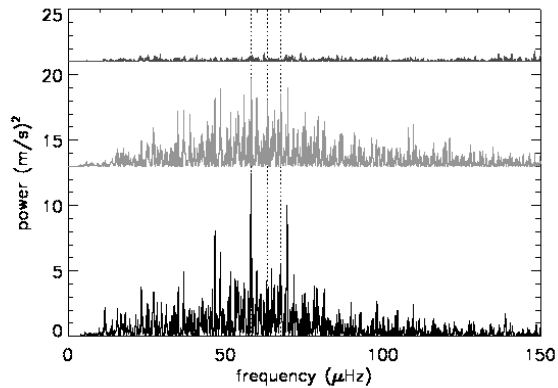}}
\caption{Power spectrum of $\epsilon$ Ophiuchi. The lower black one is obtained
by \cite{deridder2006b} from radial velocities derived with the optimum weight
method \cite{bouchy2001a}, the grey one in the middle is obtained from $\langle\mathrm{v}\rangle$ and the top one is obtained from the bisector velocity
span. For clarity the latter two are shifted. The dotted vertical lines indicate
the dominant frequencies in $\langle\mathrm{v}\rangle$. These are used in
Figure\,\protect\ref{amplphaseHD146791} to compute the amplitude across the profile.}
\label{periodogramepsoph}
\end{figure}

As mentioned earlier, \cite{deridder2006b} used the optimum weight method
\cite{bouchy2001a} to calculate the radial velocities. This is intrinsically the
most precise method, because the whole spectrum is used. With this method the
observation with the highest signal to noise ratio of each night is used as a
reference. These nightly reference points are effectively a high pass filter and
therefore the low frequencies in the power spectrum drop to a level near
zero. The resulting power spectrum of $\epsilon$ Ophiuchi is shown as the lower
black graph in Figure~\ref{periodogramepsoph}.

From this power spectrum, \cite{deridder2006b} determined two possible large
separations: 4.8 $\mu$Hz or 6.7 $\mu$Hz. Two different stellar evolution models
fit these large separations. A stellar model with a mixing length of convection
$\alpha$ = 1.6 (expressed in local pressure scale heights) and a mass of
approximately 2.8 solar masses has a large separation close to 6.7 $\mu$Hz,
assuming that the detected modes are radial.  A second stellar model with a
higher value of $\alpha$ = 1.8 and a lower mass of approximately 1.9 solar
masses has a large separation of 4.8 $\mu$Hz. The data in \cite{deridder2006b}
did not allow to discriminate between these two values of the large separations
from theoretical arguments alone.

Recently, $\epsilon$ Ophiuchi was also observed with the MOST satellite. These
continuous photometric observations, described elsewhere in this volume
\cite{barban2006}, rule out the 6.7 $\mu$Hz large separation and thus reveal the
evolutionary phase of $\epsilon$ Ophiuchi, again assuming that its detected
modes are all radial.

\subsection{Moments}
Although the radial velocities derived from the optimum weight method are
intrinsically the most accurate ones, this method is not useful for mode
identification. In order to get the best line profile analysis a cross
correlation profile of each spectrum is calculated, using a box-shaped mask
\cite{baranne1996}. This cross correlation profile has an increased signal to
noise ratio compared to a single spectral line. Such a profile can be described
by its moments and one usually considers the first three moments for mode
identification \cite{aerts1992}. The first moment $\langle \mathrm{v} \rangle$
represents the centroid velocity of the line profile, the second moment $\langle
\mathrm{v}^{2} \rangle$ the width of the line profile and the third moment
$\langle \mathrm{v}^{3} \rangle$ the skewness of the line profile. The quantity
$\langle \mathrm{v} \rangle$ is a particular measure of the radial velocity and
should therefore show similar frequency behaviour as the radial velocity derived
from the optimum weight method. The power spectrum of $\langle \mathrm{v}
\rangle$, where the average value of $\langle \mathrm{v} \rangle$ per night is
used as a reference, is shown in the middle of
Figure~\ref{periodogramepsoph}. Despite the differences in the methods to compute
a measure of the radial velocity, both power spectra reveal the same dominant
frequencies.

\subsection{Bisector Velocity Span}
A line bisector is a measure of the displacement of the centre of the red and
blue wing from the core of the spectral line at each residual flux. The bisector
velocity span is defined as the horizontal distance between the bisector
positions at fractional flux levels in the top and bottom part of the line
profile, see for instance \cite{brown1998}.

The bisector velocity span is a measure of the line profile, mostly used as a
tool to distinguish between companions and intrinsic stellar activity as sources
of the observed radial velocity variations.  The Doppler shift induced by a
companion shifts the entire spectrum, but does not change the spectral line
shapes, while oscillations and star spots do change the spectral line shapes.

No dominant frequencies are obtained from the power spectrum of the bisector
velocity span, which is plotted as the top graph in
Figure~\ref{periodogramepsoph}.  We conclude that for oscillations with low amplitudes, the
bisector velocity span is not a useful diagnostic to distinguish between
planetary companions and intrinsic variations in a star. This result is
consistent with the results recently found by \cite{dall2006}.

\subsection{Mode identification}
%
%
\begin{figure*}
\begin{minipage}{5.6cm}
\centering
\includegraphics[width=5.6cm]{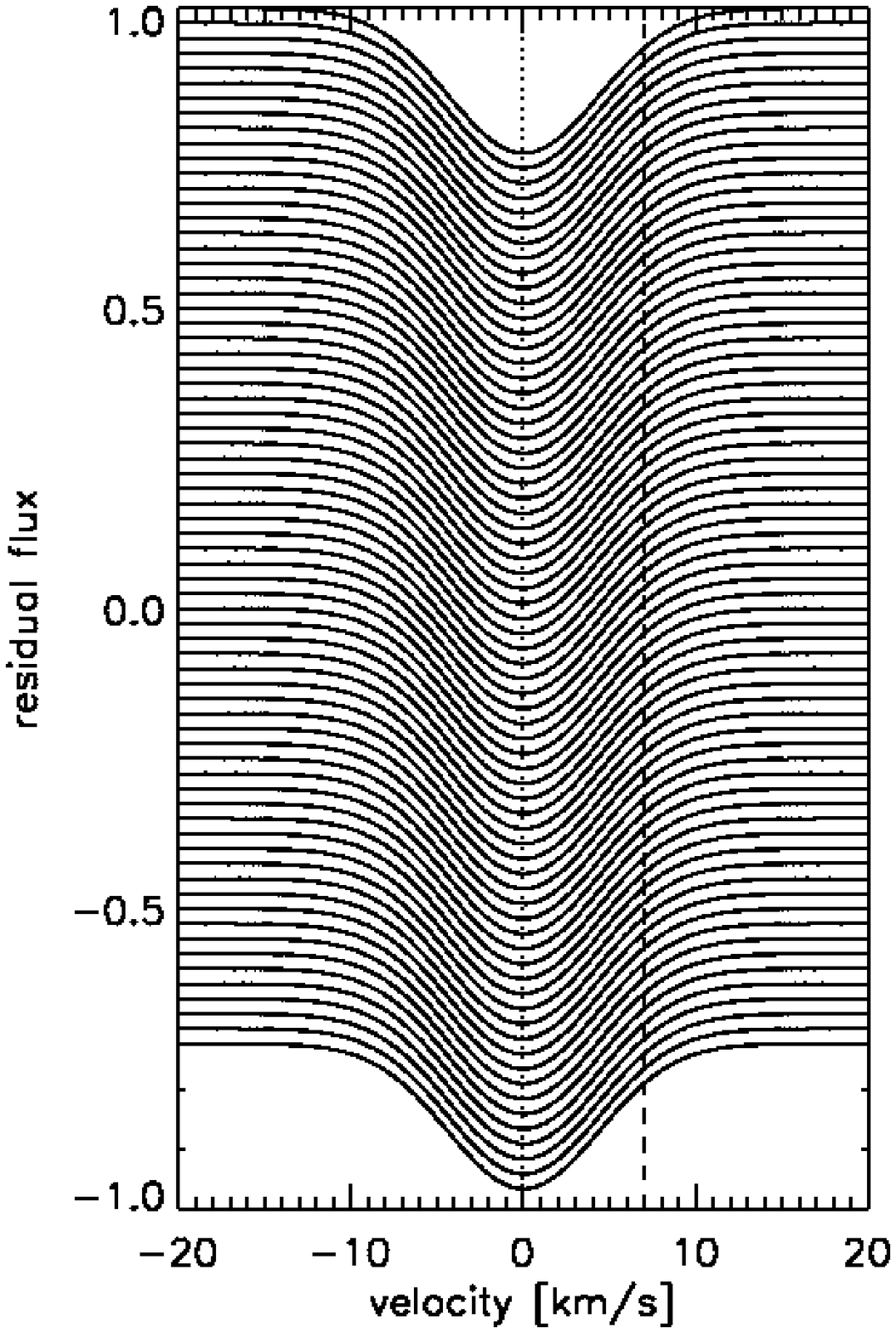}
\end{minipage}
\hfill
\begin{minipage}{5.6cm}
\begin{minipage}{5.6cm}
\centering
\includegraphics[width=5.6cm]{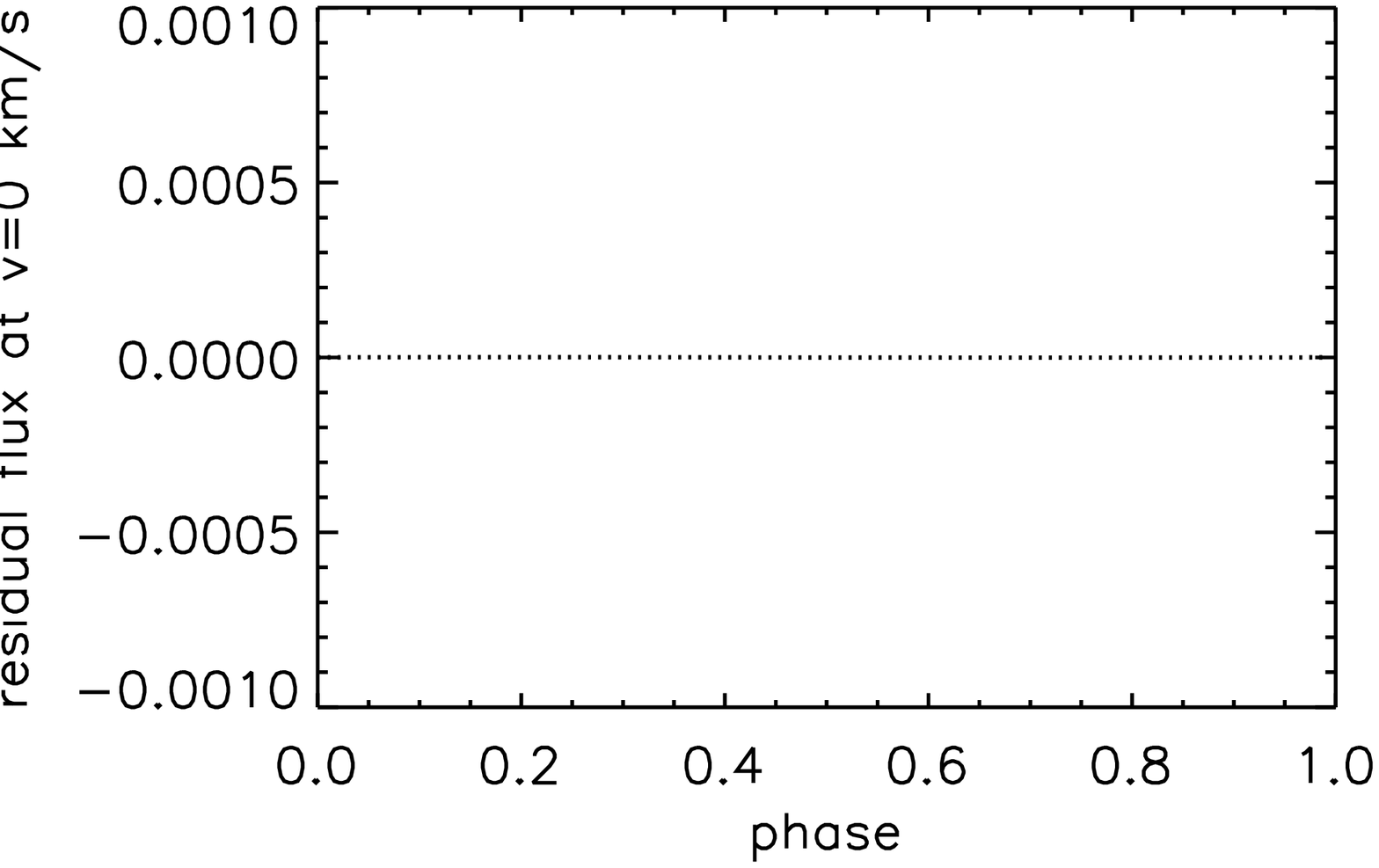}
\end{minipage}
\hfill
\begin{minipage}{5.6cm}
\centering
\includegraphics[width=5.5cm]{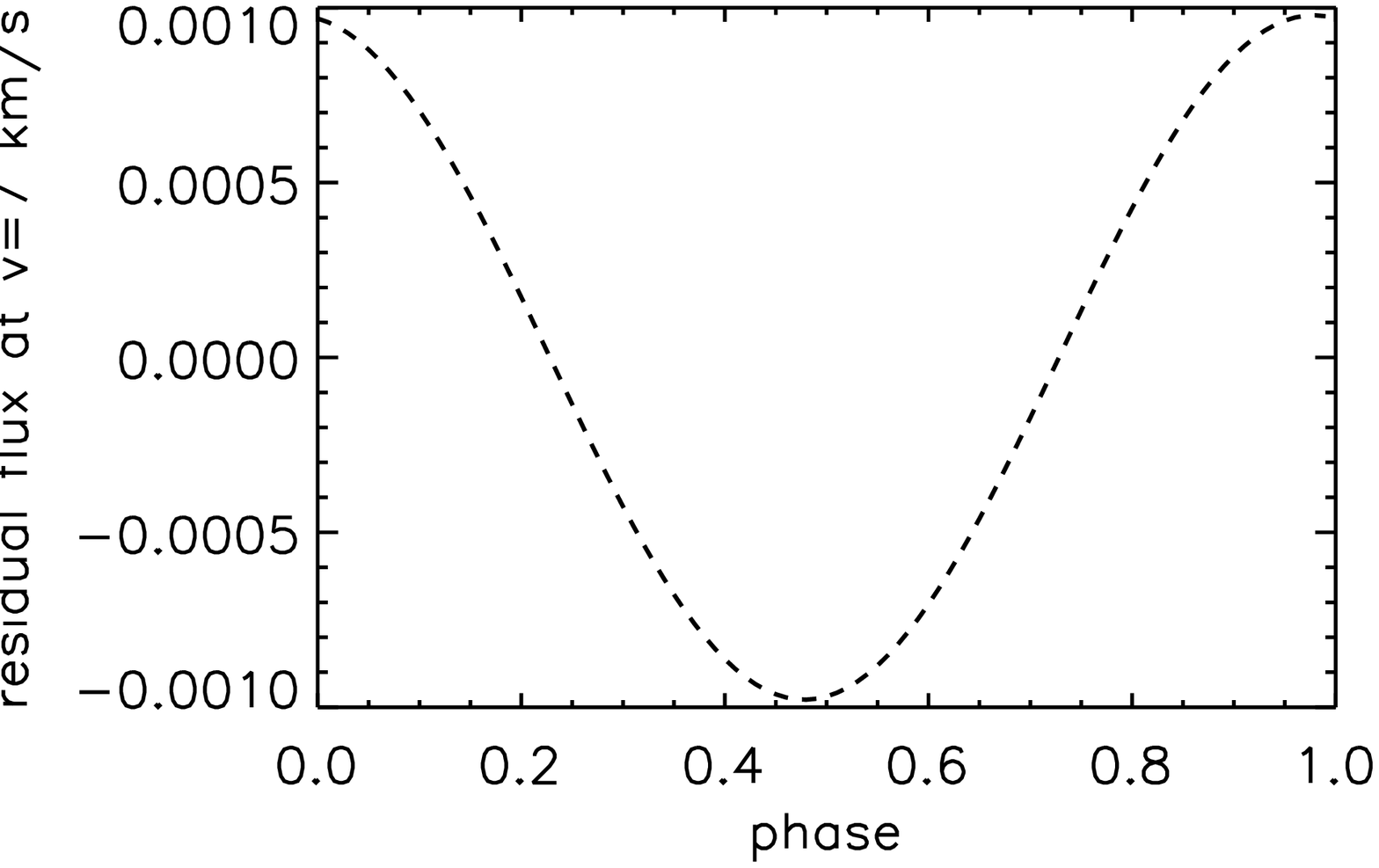}
\end{minipage}
\hfill
\end{minipage}
\begin{minipage}{5.6cm}
\begin{minipage}{5.6cm}
\centering
\includegraphics[width=5.6cm]{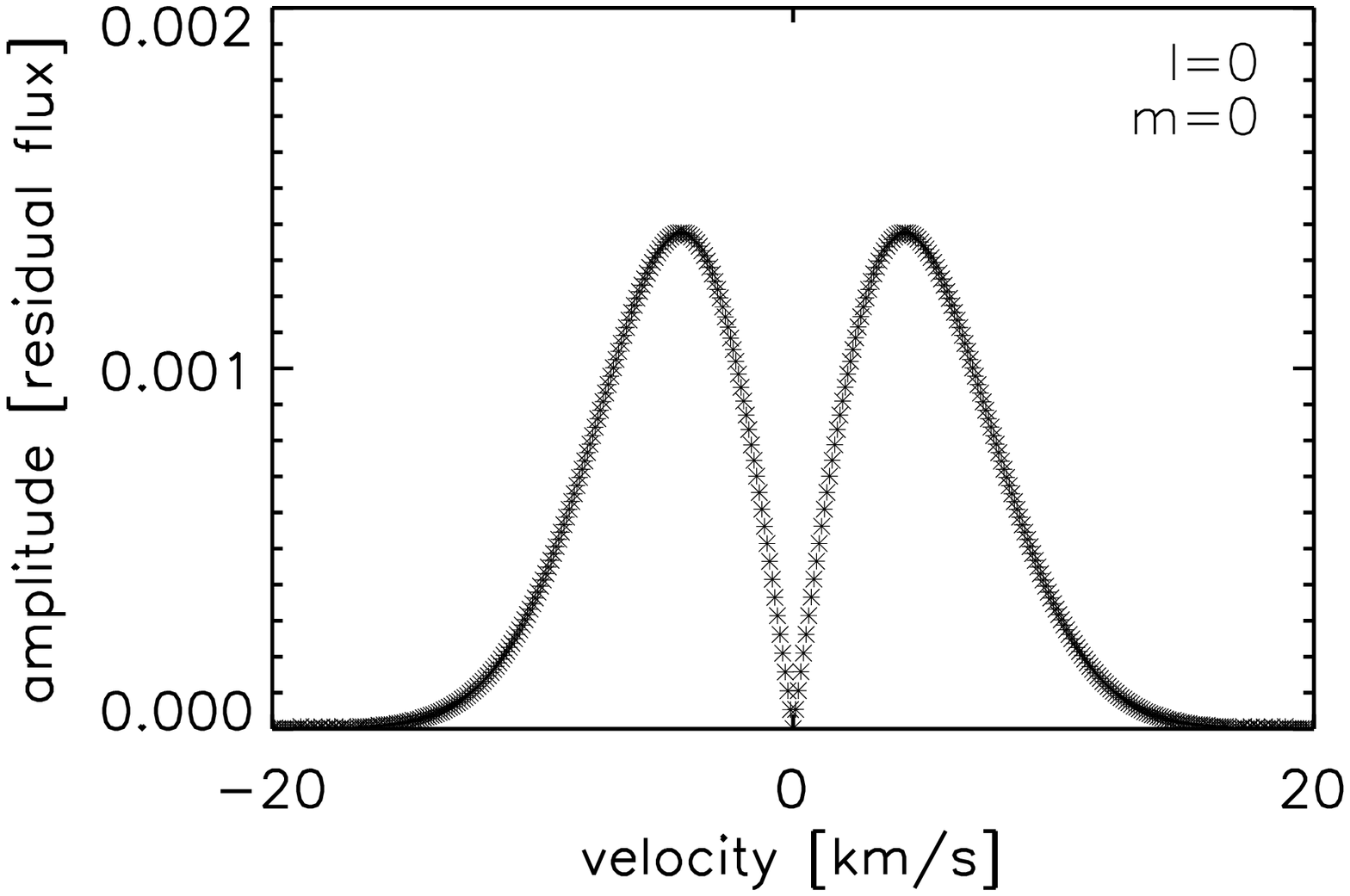}
\end{minipage}
\hfill
\begin{minipage}{5.6cm}
\centering
\includegraphics[width=5.5cm]{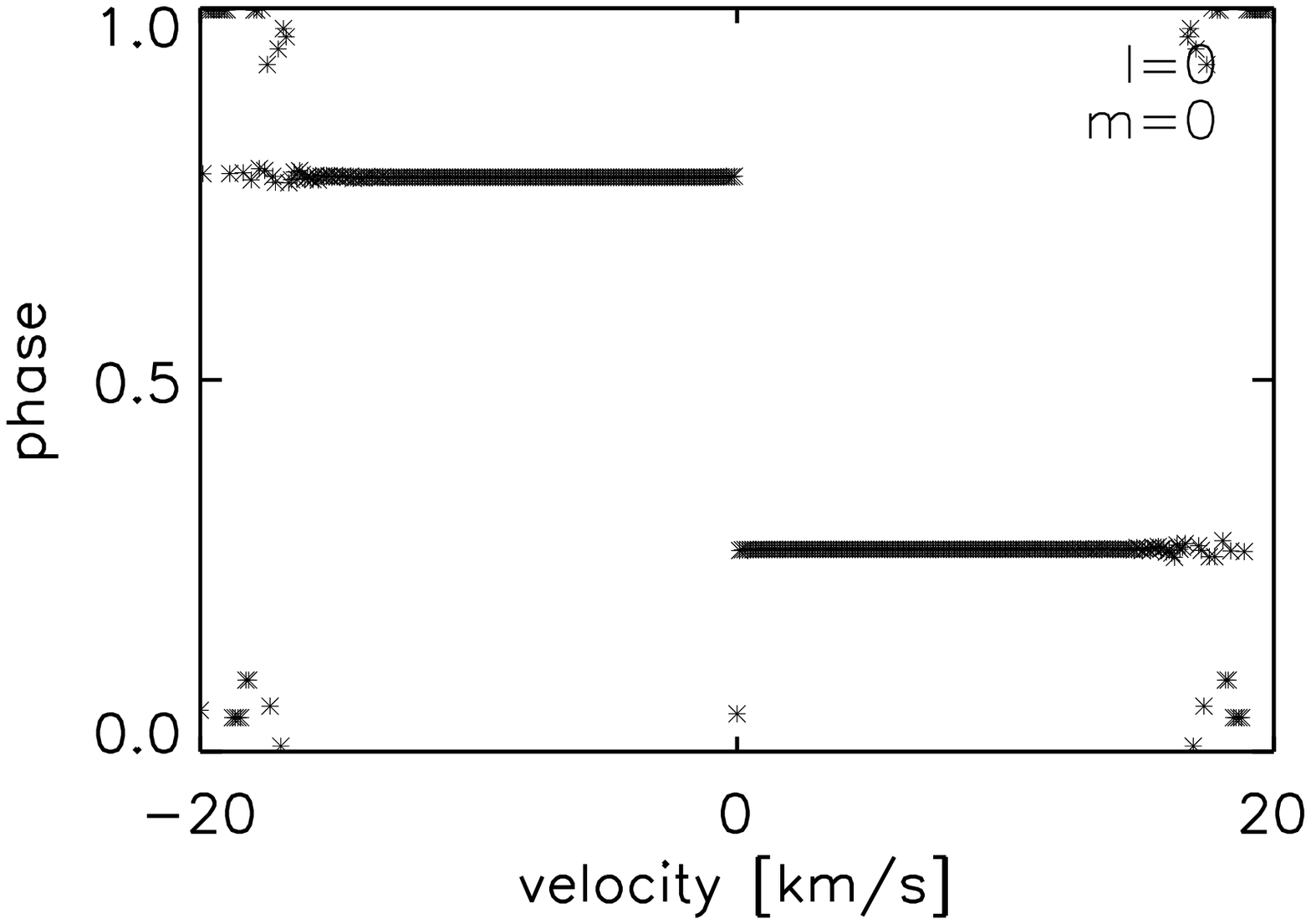}
\end{minipage}
\hfill
\end{minipage}
\caption{Schematic representation of the amplitude distribution across the
profile for simulated data with $(\ell,m)=(0,0)$, an amplitude of the pulsation velocity of 0.04 km\,s$^{-1}$, an
inclination angle of $35^{\circ}$, and intrinsic line width of 4 km\,s$^{-1}$ and a $\upsilon \mathrm{sin}i$ of 3.5 km\,s$^{-1}$.
Left: Profiles obtained at different times are shown with an arbitrary
flux shift. The dashed and dotted
lines indicate the two velocity values at which the
harmonic fits shown in the two middle panels are obtained. Middle top: harmonic fit at
the centre of the profiles. Middle bottom: harmonic fit at a wing of the profiles. Right top: amplitude across the whole profile. Right bottom: phase across the whole profile.}
\label{ampphasescheml0m0}
\end{figure*}

\begin{figure*}
\begin{minipage}{5.6cm}
\centering
\includegraphics[width=5.6cm]{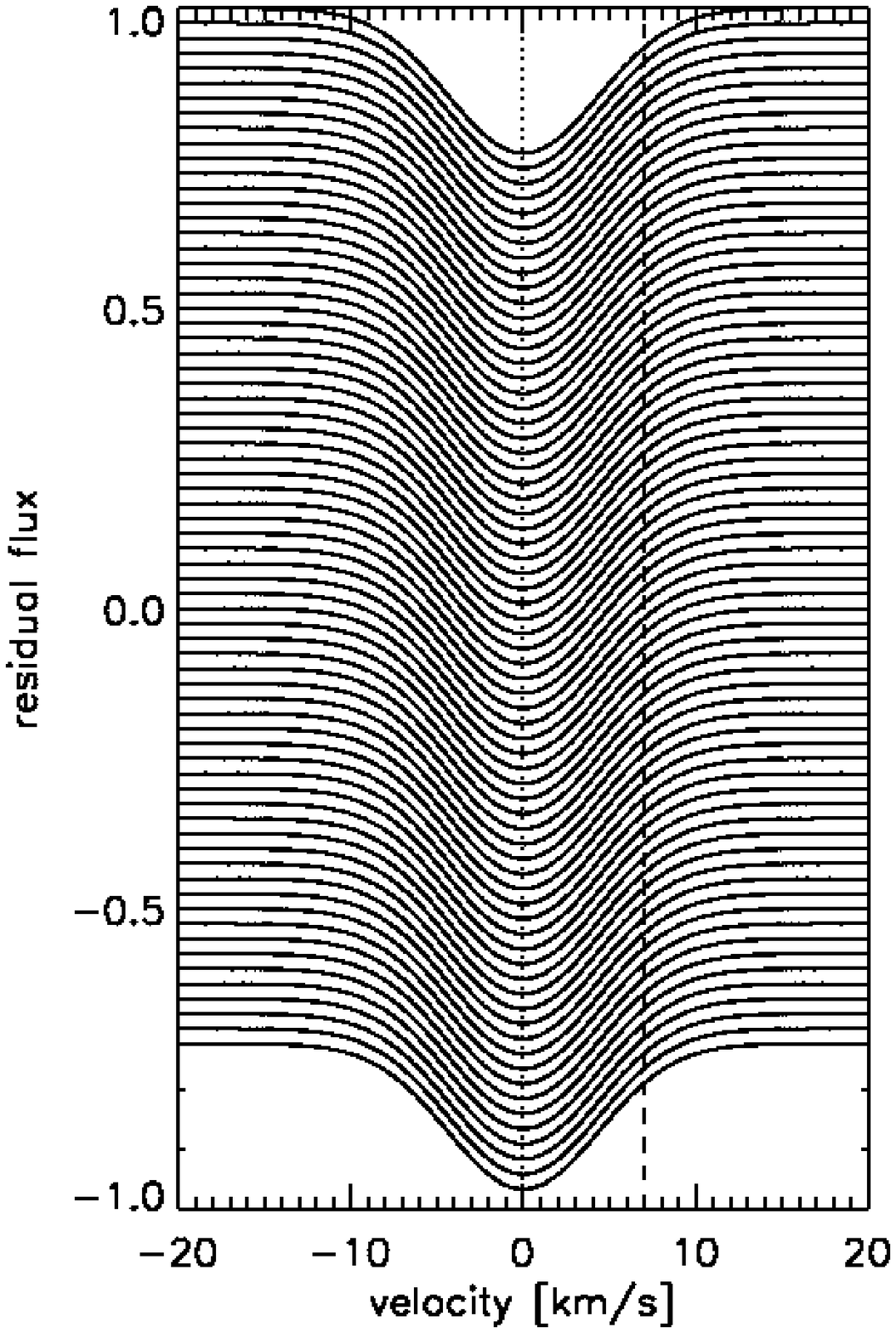}
\end{minipage}
\hfill
\begin{minipage}{5.6cm}
\begin{minipage}{5.6cm}
\centering
\includegraphics[width=5.6cm]{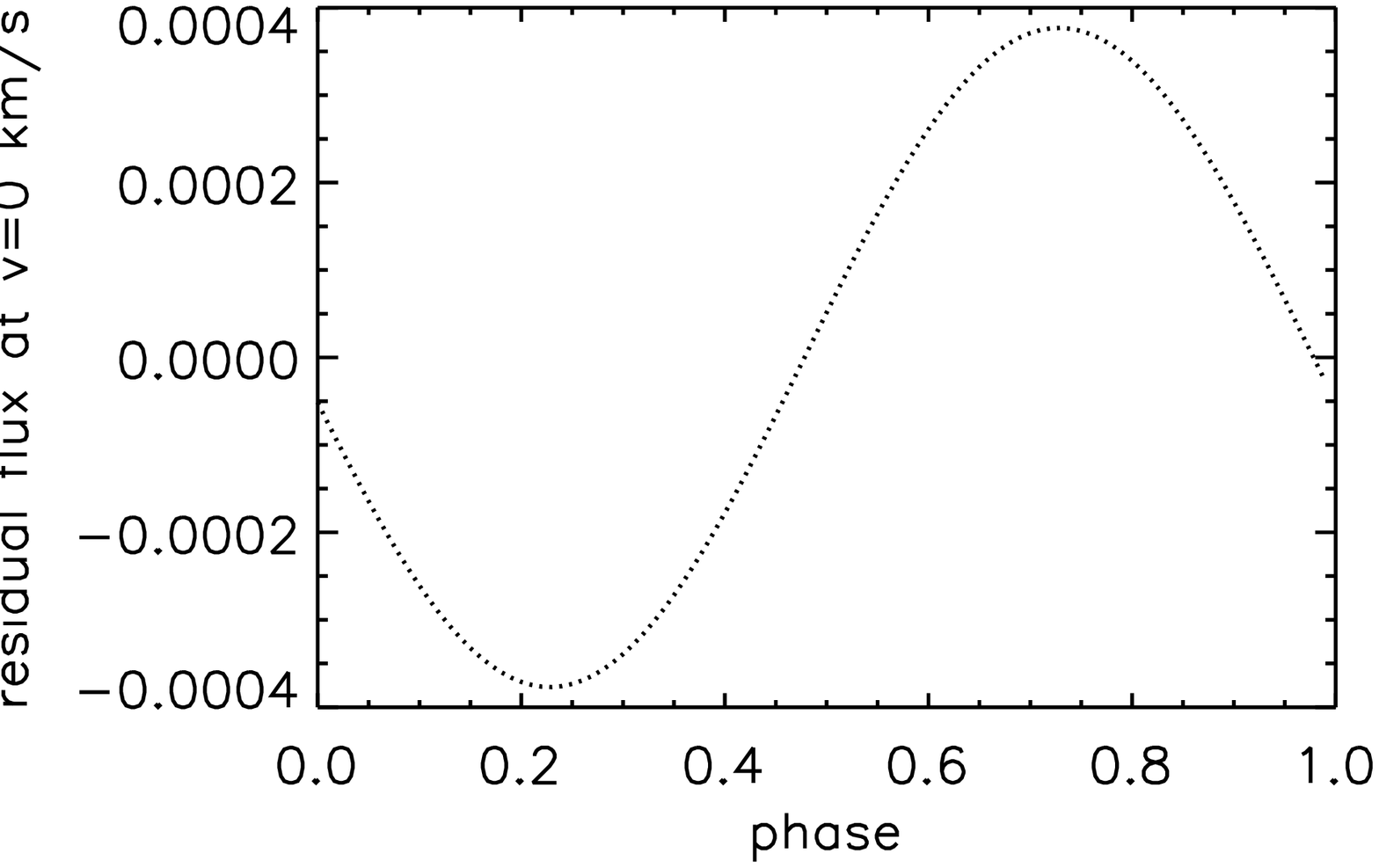}
\end{minipage}
\hfill
\begin{minipage}{5.6cm}
\centering
\includegraphics[width=5.5cm]{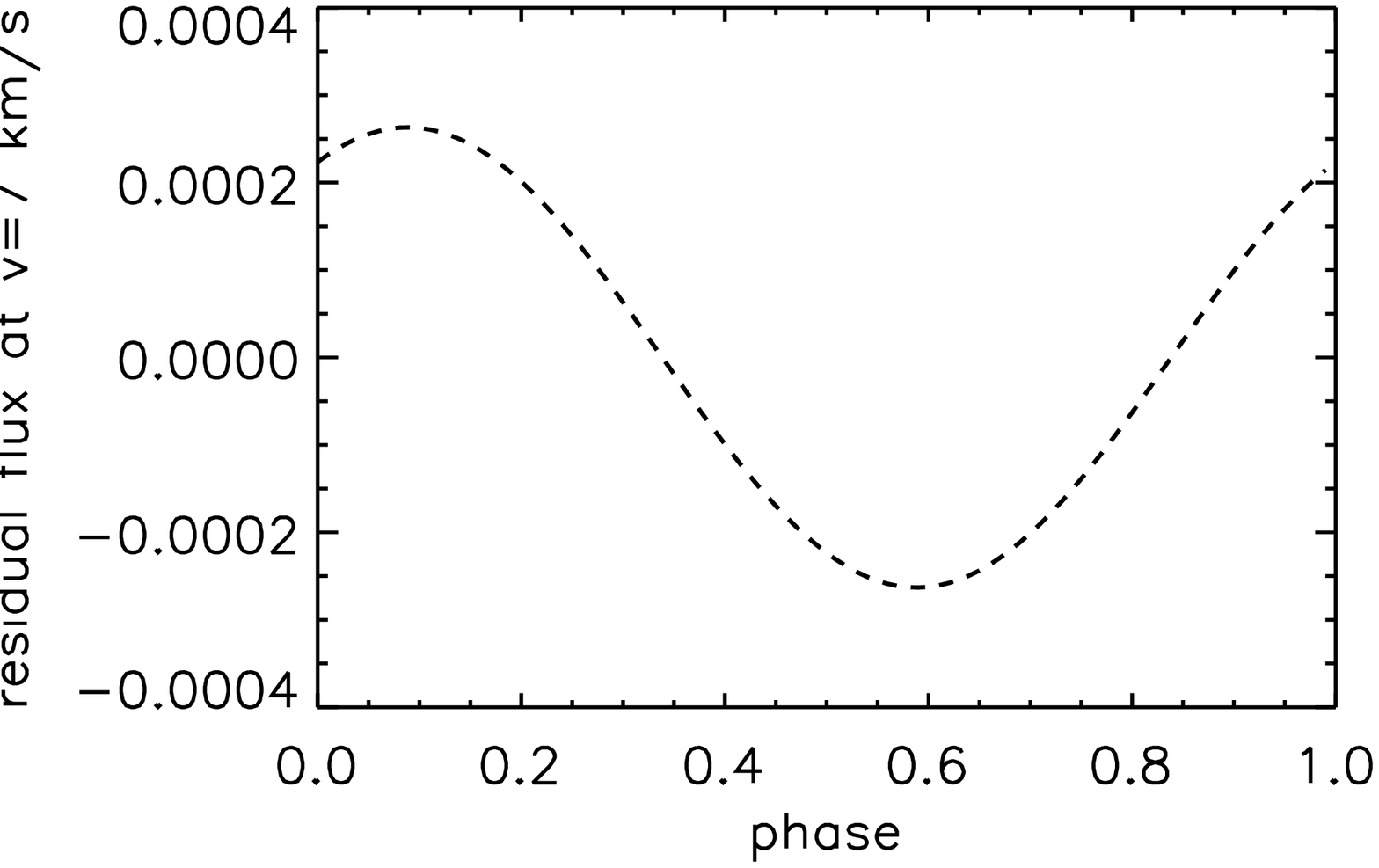}
\end{minipage}
\hfill
\end{minipage}
\begin{minipage}{5.6cm}
\begin{minipage}{5.6cm}
\centering
\includegraphics[width=5.6cm]{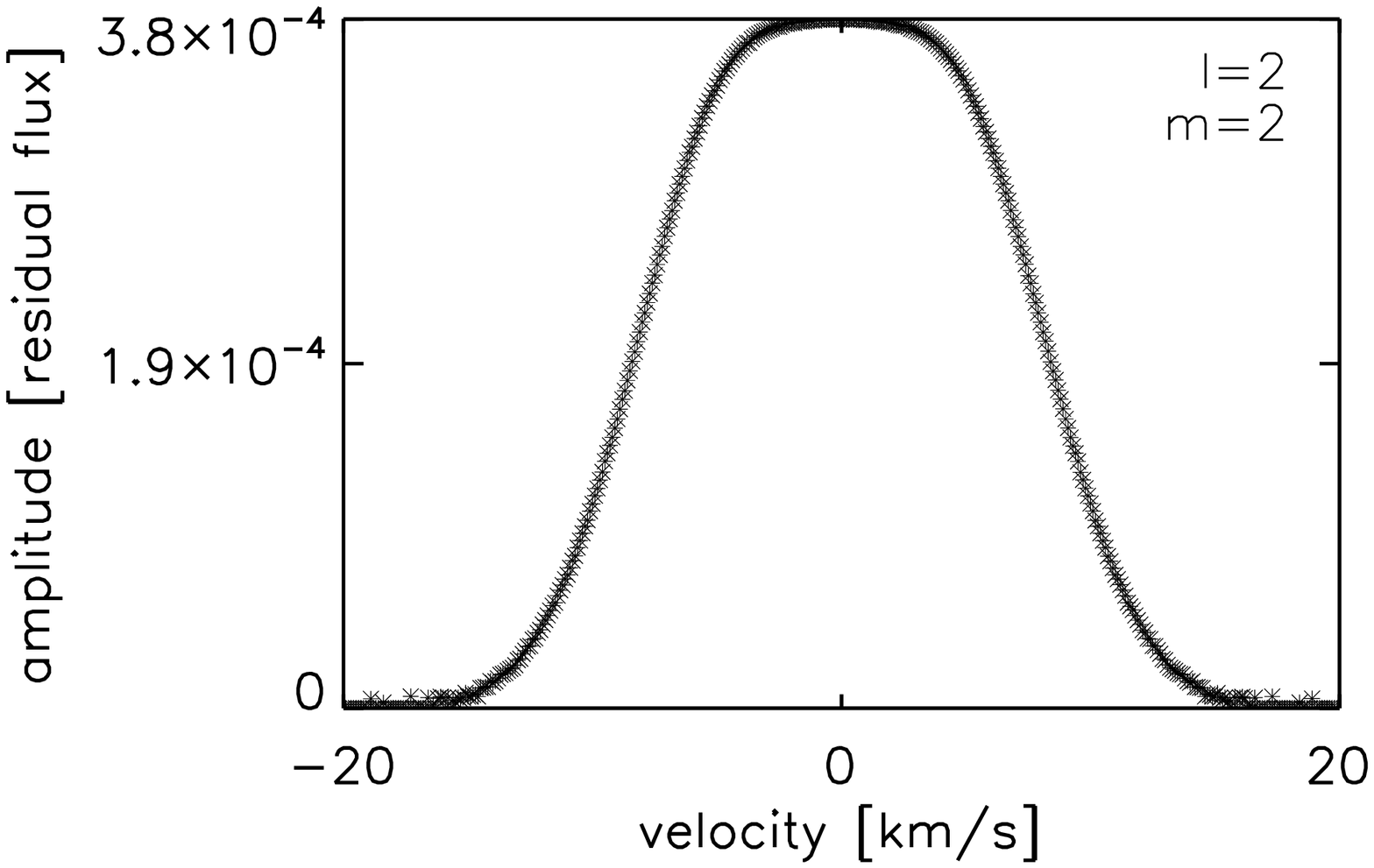}
\end{minipage}
\hfill
\begin{minipage}{5.6cm}
\centering
\includegraphics[width=5.5cm]{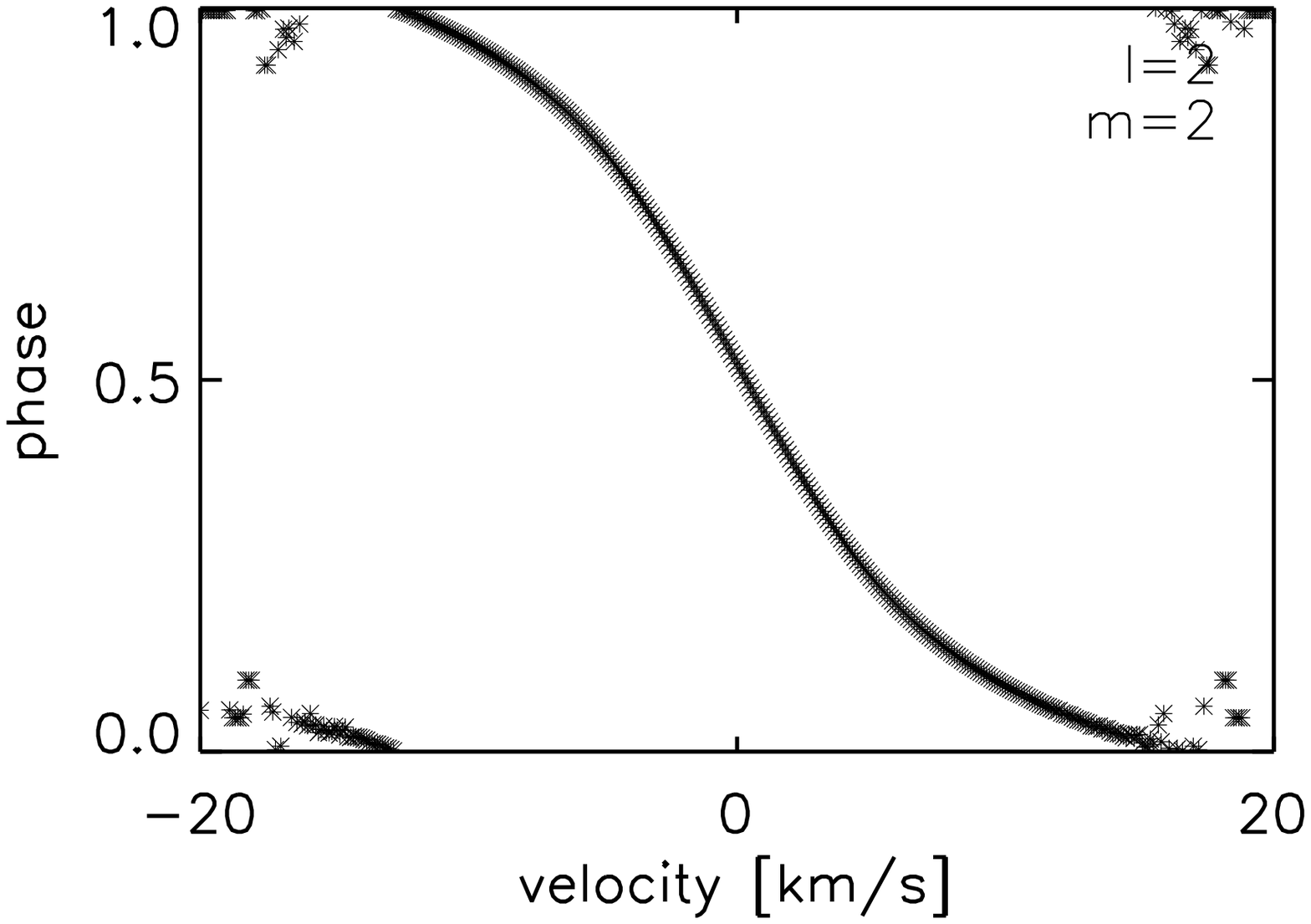}
\end{minipage}
\hfill
\end{minipage}
\caption{Schematic representation of the amplitude distribution across the
profile for simulated data with $(\ell,m)=(2,2)$, an amplitude of the pulsation velocity of 0.04 km\,s$^{-1}$, an
inclination angle of $35^{\circ}$, and intrinsic line width of 4 km\,s$^{-1}$ and a $\upsilon \mathrm{sin}i$ of 3.5 km\,s$^{-1}$.
Left: Profiles obtained at different times are shown with an arbitrary
flux shift. The dashed and dotted
lines indicate the two velocity values at which the
harmonic fits shown in the two middle panels are obtained. Middle top: harmonic fit at
the centre of the profiles. Middle bottom: harmonic fit at a wing of the profiles. Right top: amplitude across the whole profile. Right bottom: phase across the whole profile.}
\label{ampphasescheml2m2}
\end{figure*}

\begin{figure*}
\begin{minipage}{5.6cm}
\centering
\includegraphics[width=5.5cm]{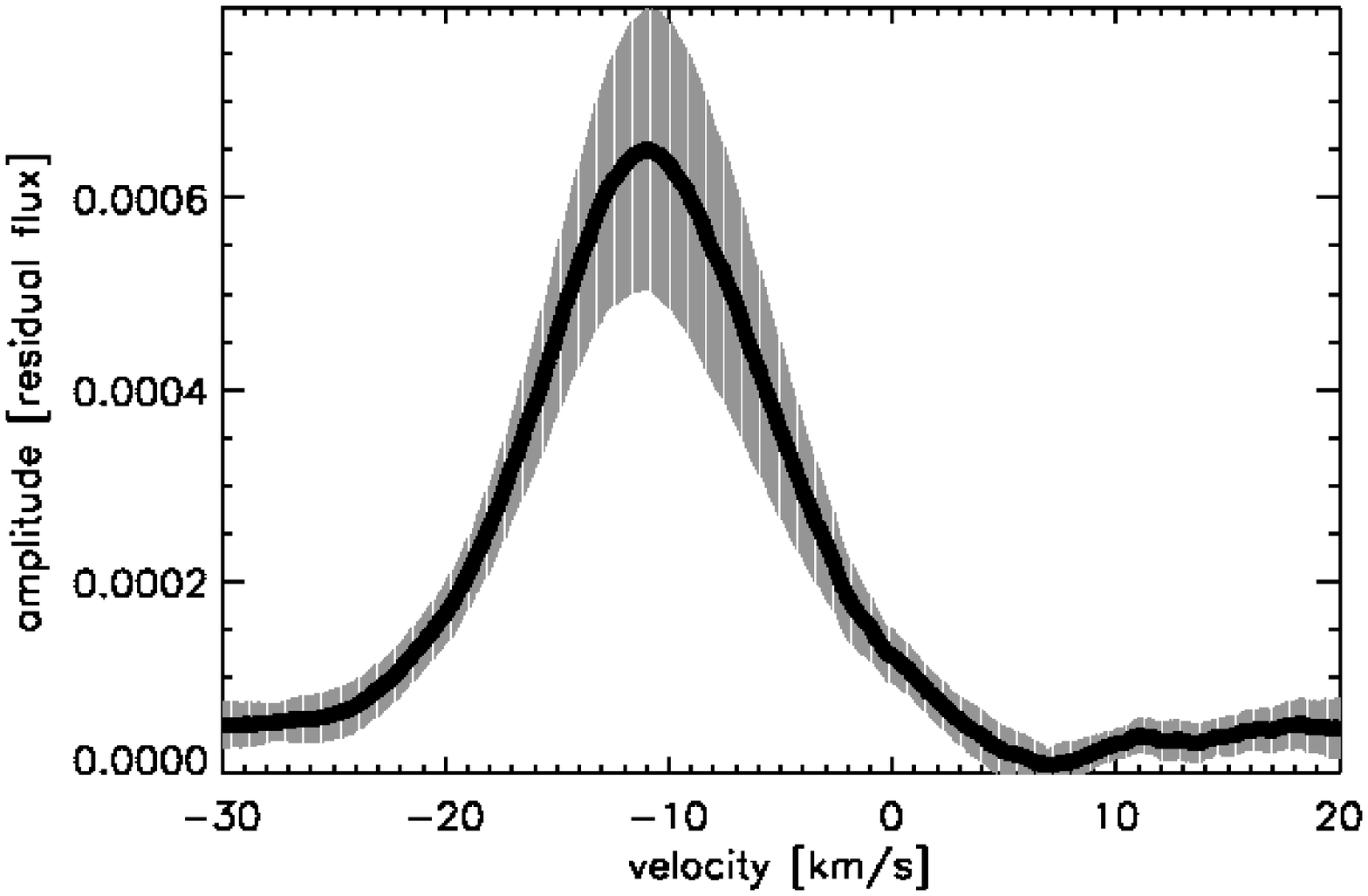}
\end{minipage}
\hfill
\begin{minipage}{5.6cm}
\centering
\includegraphics[width=5.5cm]{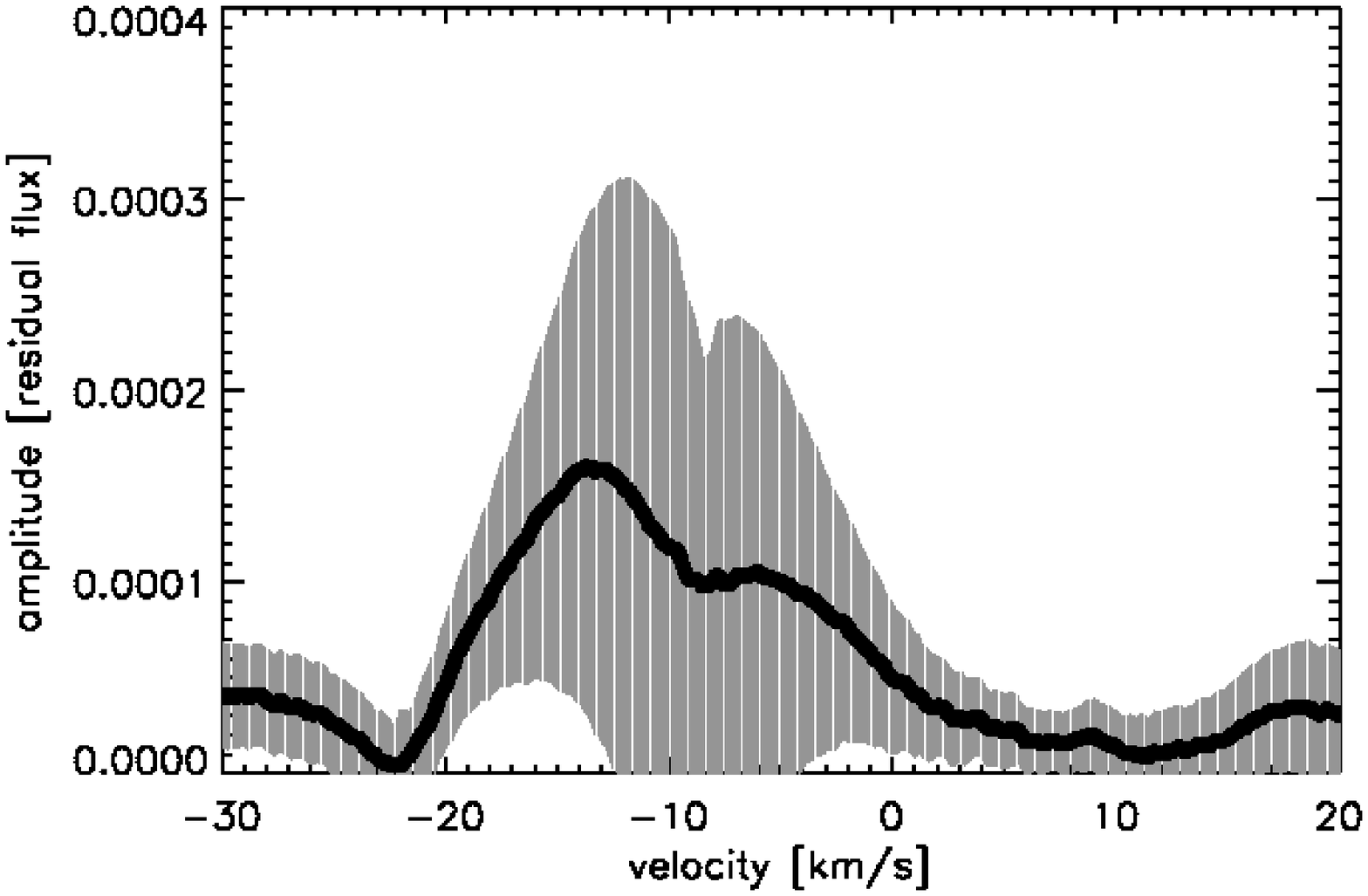}
\end{minipage}
\hfill
\begin{minipage}{5.6cm}
\centering
\includegraphics[width=5.5cm]{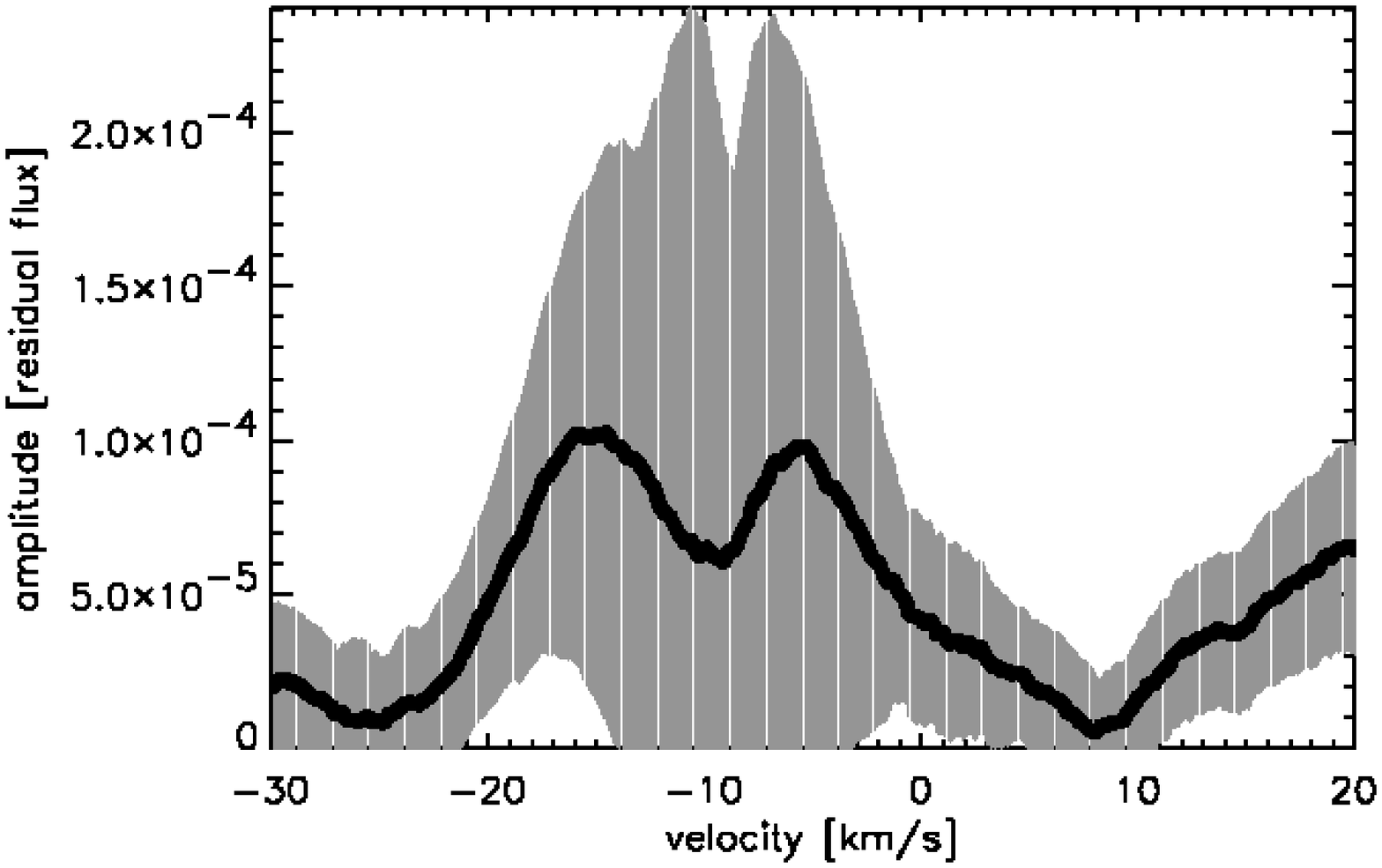}
\end{minipage}
\hfill
\begin{minipage}{5.6cm}
\centering
\includegraphics[width=5.5cm]{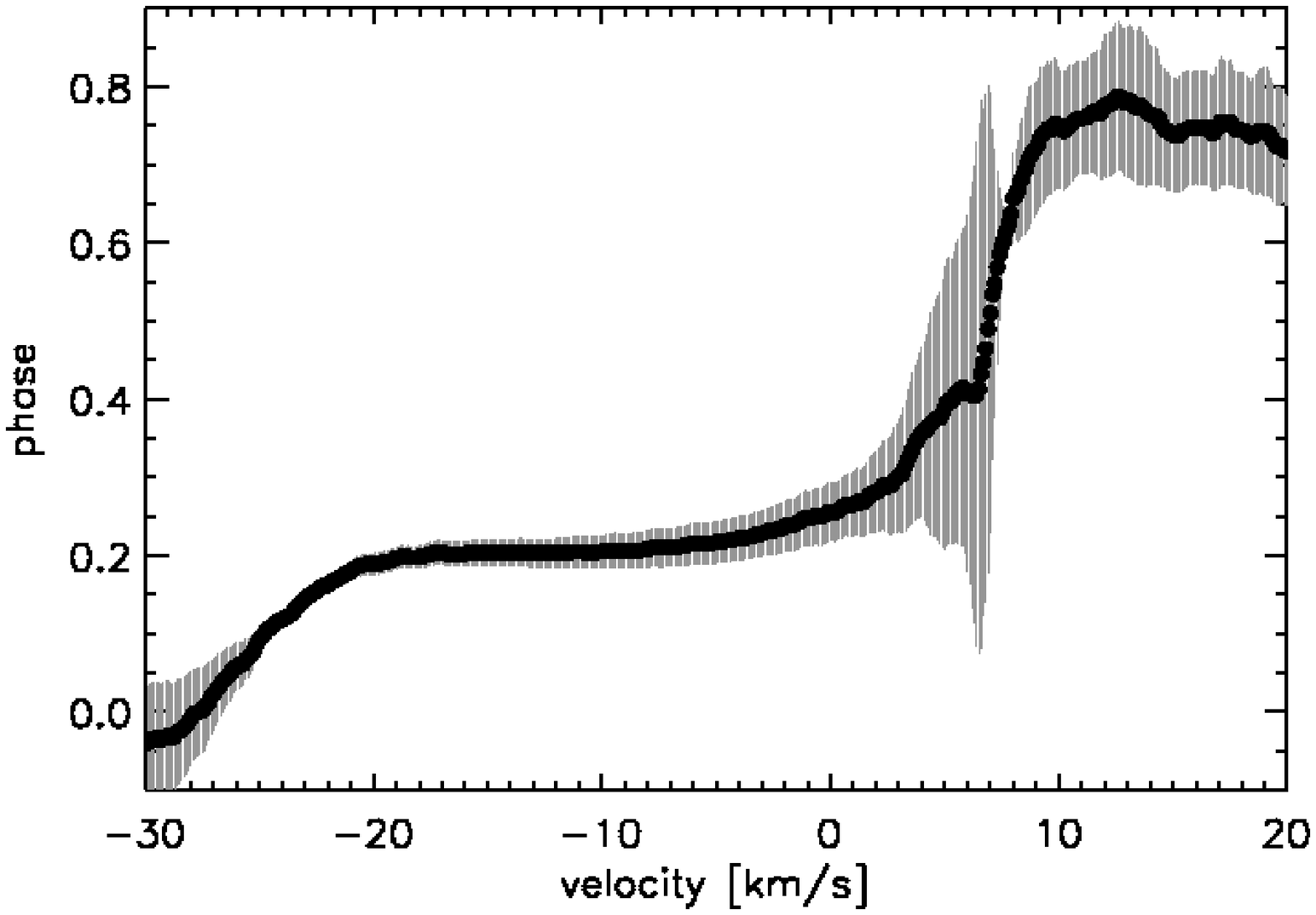}
\end{minipage}
\hfill
\begin{minipage}{5.6cm}
\centering
\includegraphics[width=5.5cm]{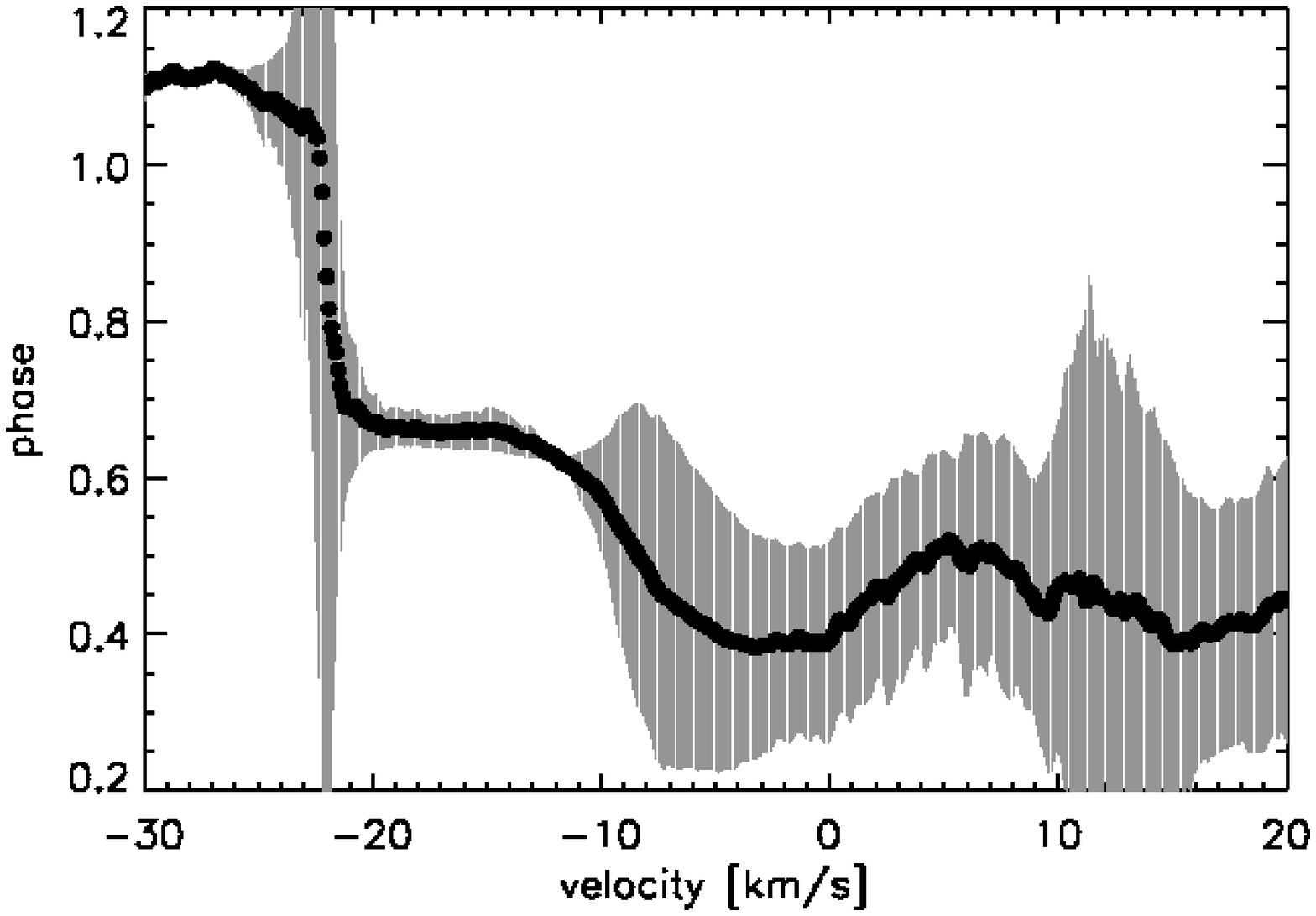}
\end{minipage}
\hfill
\begin{minipage}{5.6cm}
\centering
\includegraphics[width=5.5cm]{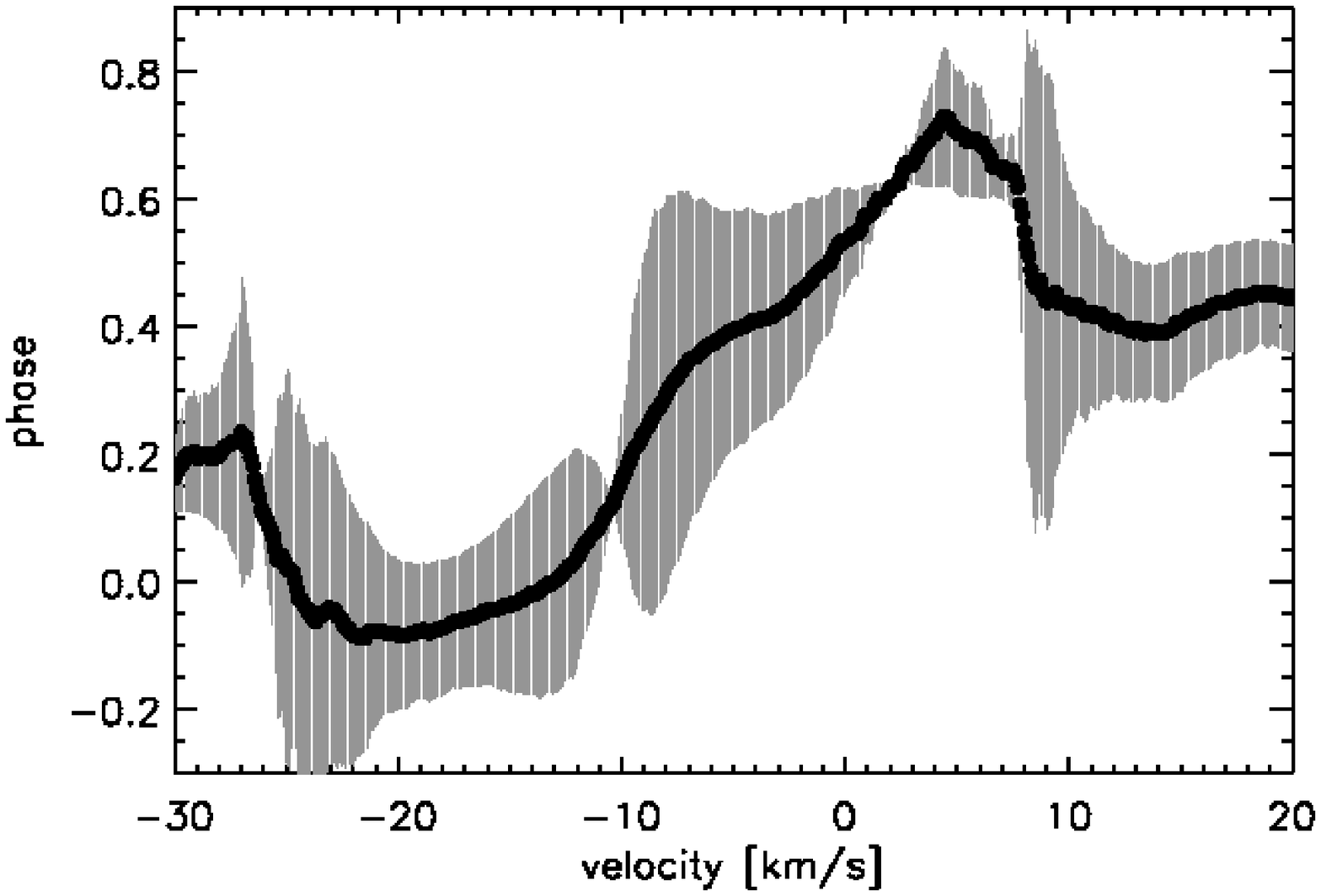}
\end{minipage}
\caption{Amplitude (top) and phase (bottom) distributions (thick lines) and their uncertainty
  (indicated in grey) as a function of velocity across the line profile of
  $\epsilon$ Ophiuchi for three significant frequencies obtained from $\langle
  \mathrm{v} \rangle$: $\nu_{\langle \mathrm{v} \rangle}=5.03$ ~c/d
  ($58.2\mu$Hz) (left) $\nu_{\langle \mathrm{v} \rangle}=5.46$ ~c/d
  ($63.2\mu$Hz) (middle) and $\nu_{\langle \mathrm{v} \rangle}=5.83$ ~c/d
  ($67.5\mu$Hz) (right). The mean radial velocity of the star, is found to be
  approximately $-9.4$~km/s.}
\label{amplphaseHD146791}
\end{figure*}

In order to characterise the wavenumbers ($\ell,m$) of the oscillation modes,
the amplitudes across the line profile are compared with the ones obtained from
simulations. These amplitudes are determined by fitting a harmonic function,
with the dominant frequency of $\langle \mathrm{v} \rangle$, to the flux values
at each velocity pixel of the time series of spectra \cite{schrijvers1997}.
This is shown schematically in Figure~\ref{ampphasescheml0m0} for $(\ell,m)=(0,0)$, and in Figure~\ref{ampphasescheml2m2} for $(\ell,m)=(2,2)$. The left panel shows
profiles obtained at different times with an arbitrary flux shift. The dotted / dashed
lines indicate two examples of velocity points at which a harmonic function is
fitted.  A harmonic fit at the centre of the profiles is shown in the top middle
panel of Figures~\ref{ampphasescheml0m0} and \ref{ampphasescheml2m2} as a function of phase. A harmonic fit
through a wing of the profiles is shown in the bottom middle panel of
Figures~\ref{ampphasescheml0m0} and \ref{ampphasescheml2m2}. A difference between the amplitudes of the harmonic
fit at the different velocity points is clearly seen. The behaviour of the
amplitude across the whole profile is shown in the top right panel, while the phase across the whole profile is shown in the bottom right panel. This
diagnostic is suitable to be compared with a similar one derived from the data.

The observed amplitude and phase distributions of $\epsilon$ Ophiuchi for its three
dominant frequencies are shown in Figure~\ref{amplphaseHD146791}, as a function
of velocity across the line profile. If only radial modes would be dominant in
$\epsilon$ Ophiuchi's line profile variations, then all three amplitude distributions
should have the same shape. However, the detected shapes of these amplitude diagrams clearly differ
for the three frequencies. Moreover, the very characteristic amplitude
shape for an axisymmetric $(m=0)$ mode is not recovered for the dominant mode of
the star. This implies that at least one non-radial mode is detected in the line
profiles of $\epsilon$ Ophiuchi.

\section{Simulations of line profiles}
\begin{figure*}
\begin{minipage}{8.6cm}
\begin{minipage}{4.25cm}
\centering
\includegraphics[width=4.25cm]{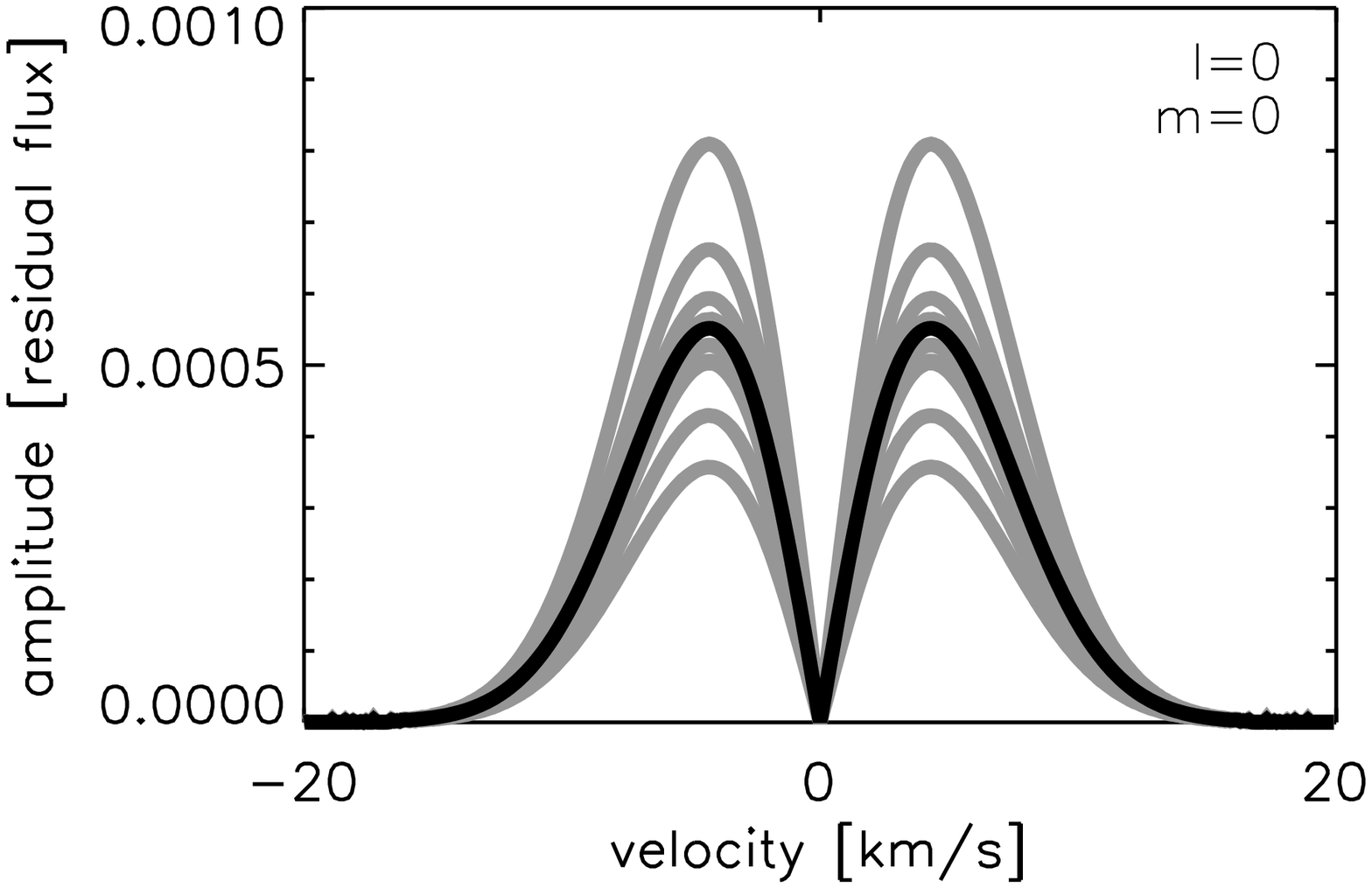}
\end{minipage}
\hfill
\begin{minipage}{4.25cm}
\centering
\includegraphics[width=4.25cm]{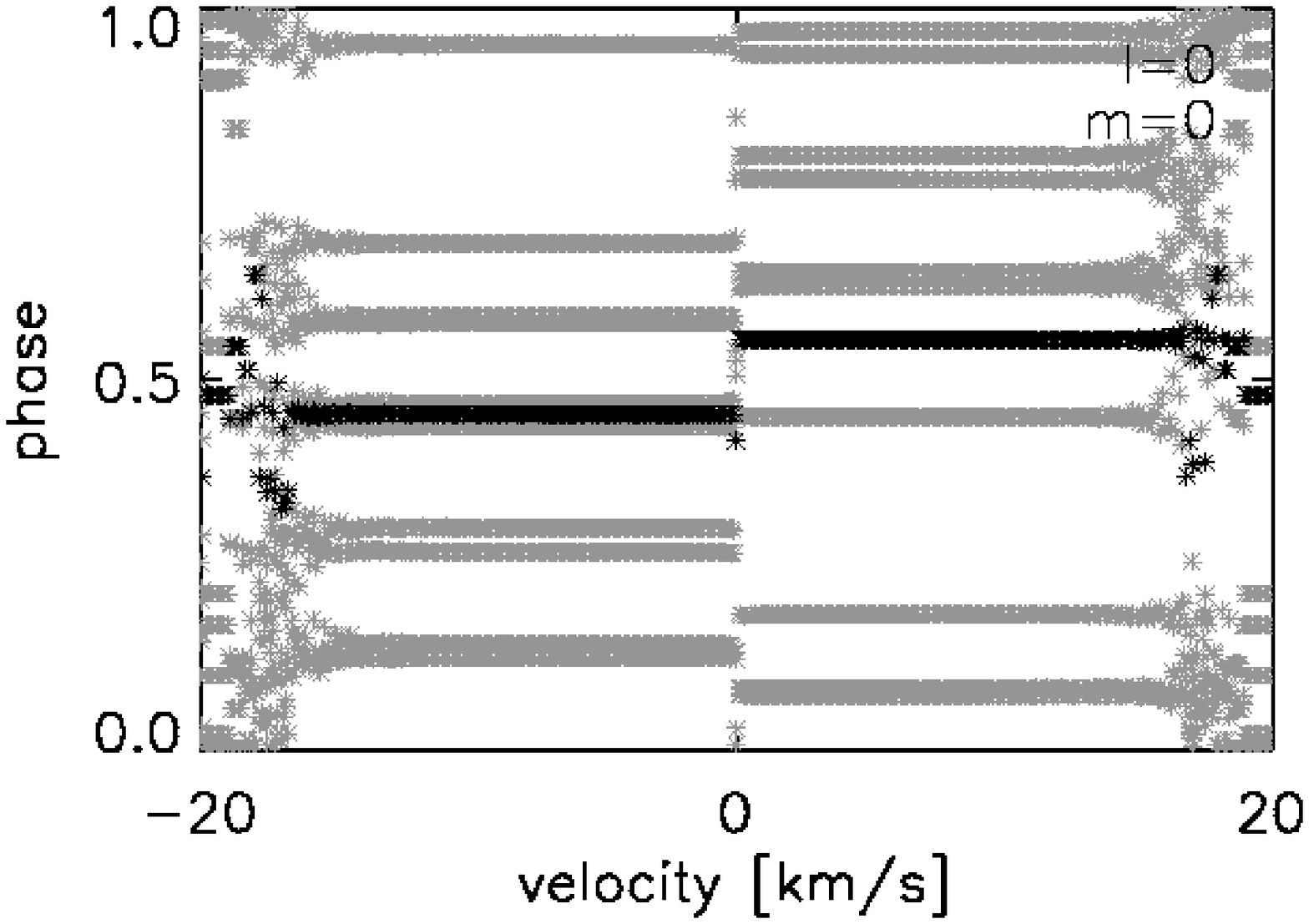}
\end{minipage}
\end{minipage}
\begin{minipage}{8.6cm}
\begin{minipage}{4.25cm}
\centering
\includegraphics[width=4.25cm]{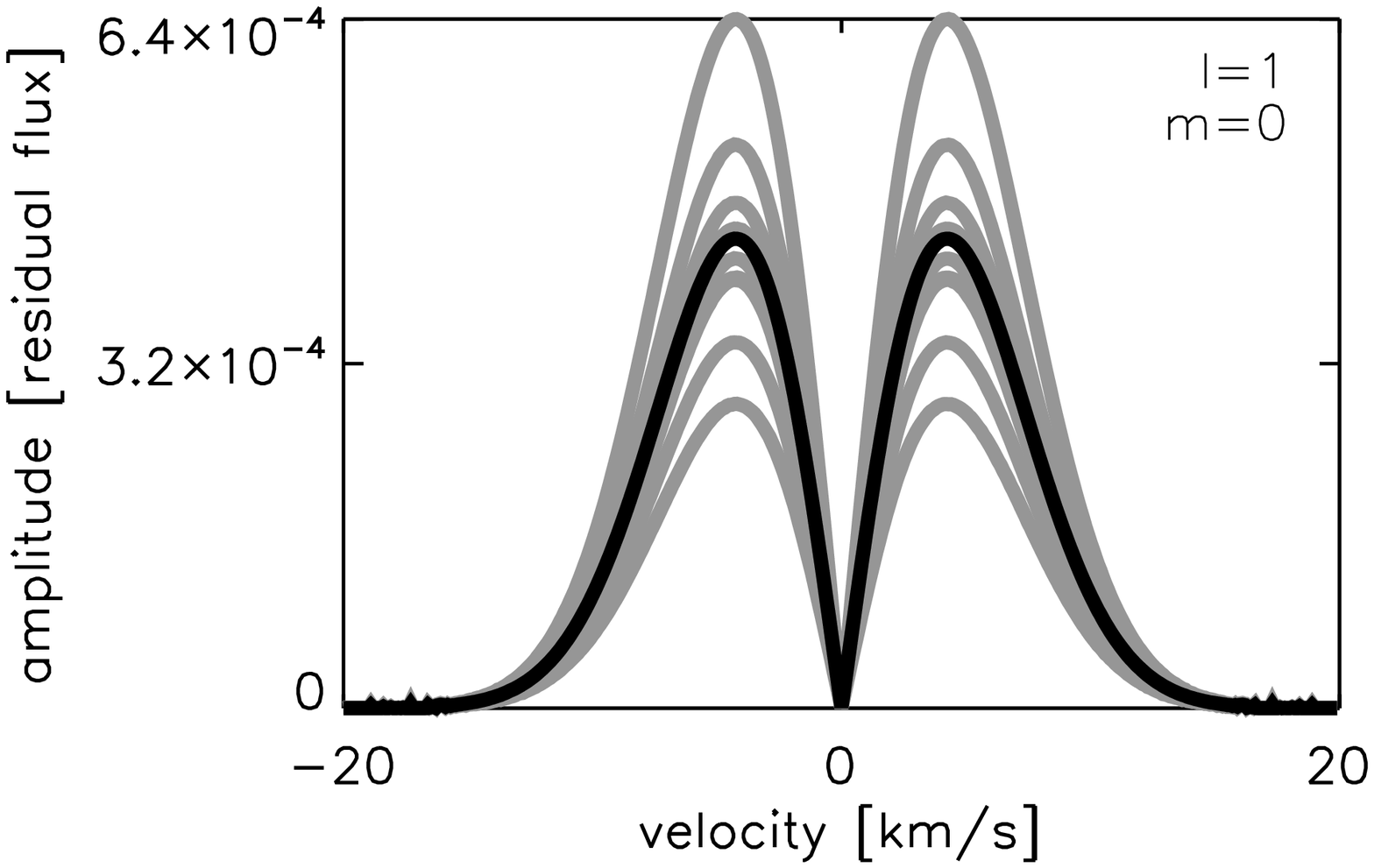}
\end{minipage}
\hfill
\begin{minipage}{4.25cm}
\centering
\includegraphics[width=4.25cm]{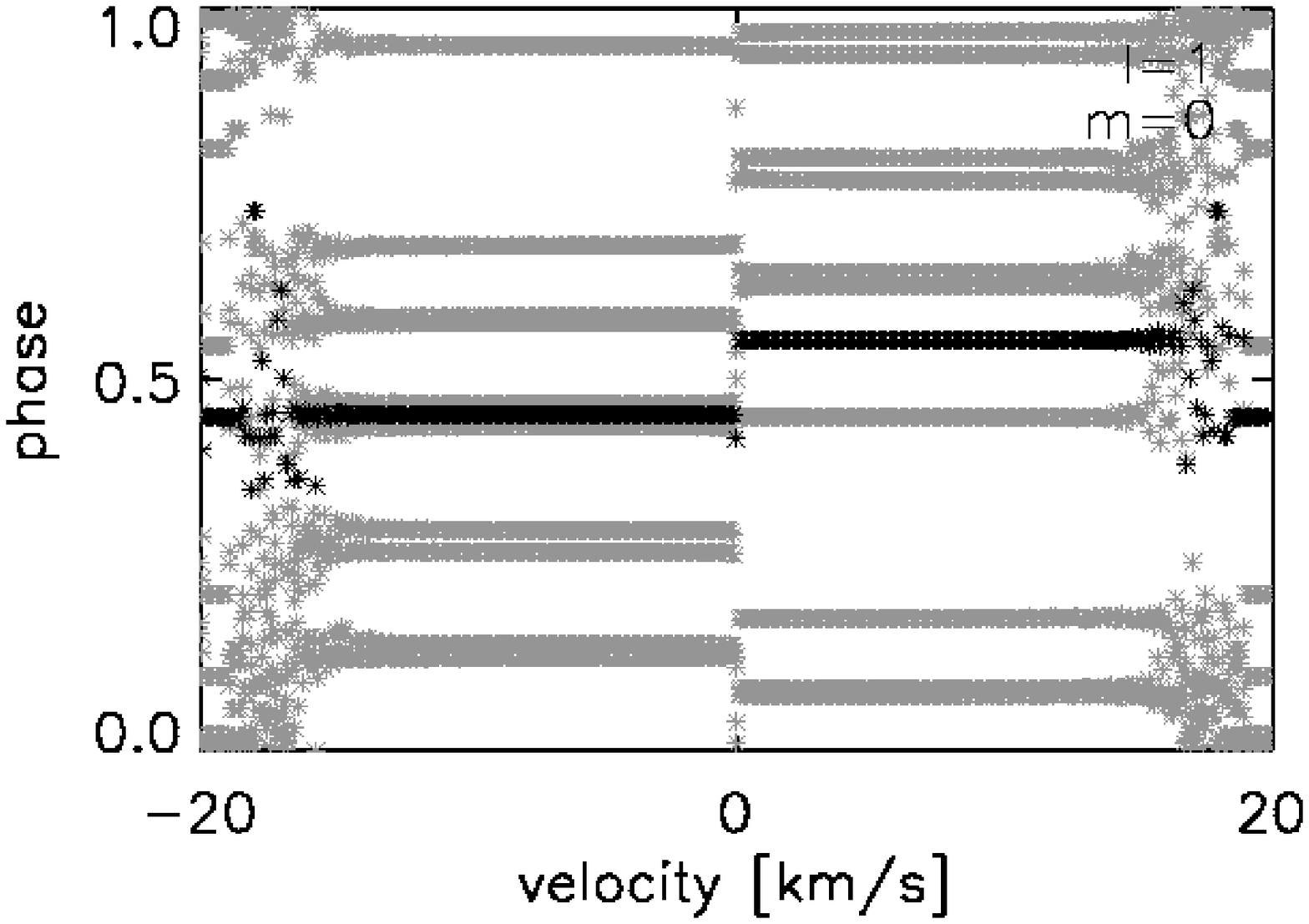}
\end{minipage}
\end{minipage}
\begin{minipage}{8.6cm}
\begin{minipage}{4.2cm}
\centering
\includegraphics[width=4.2cm]{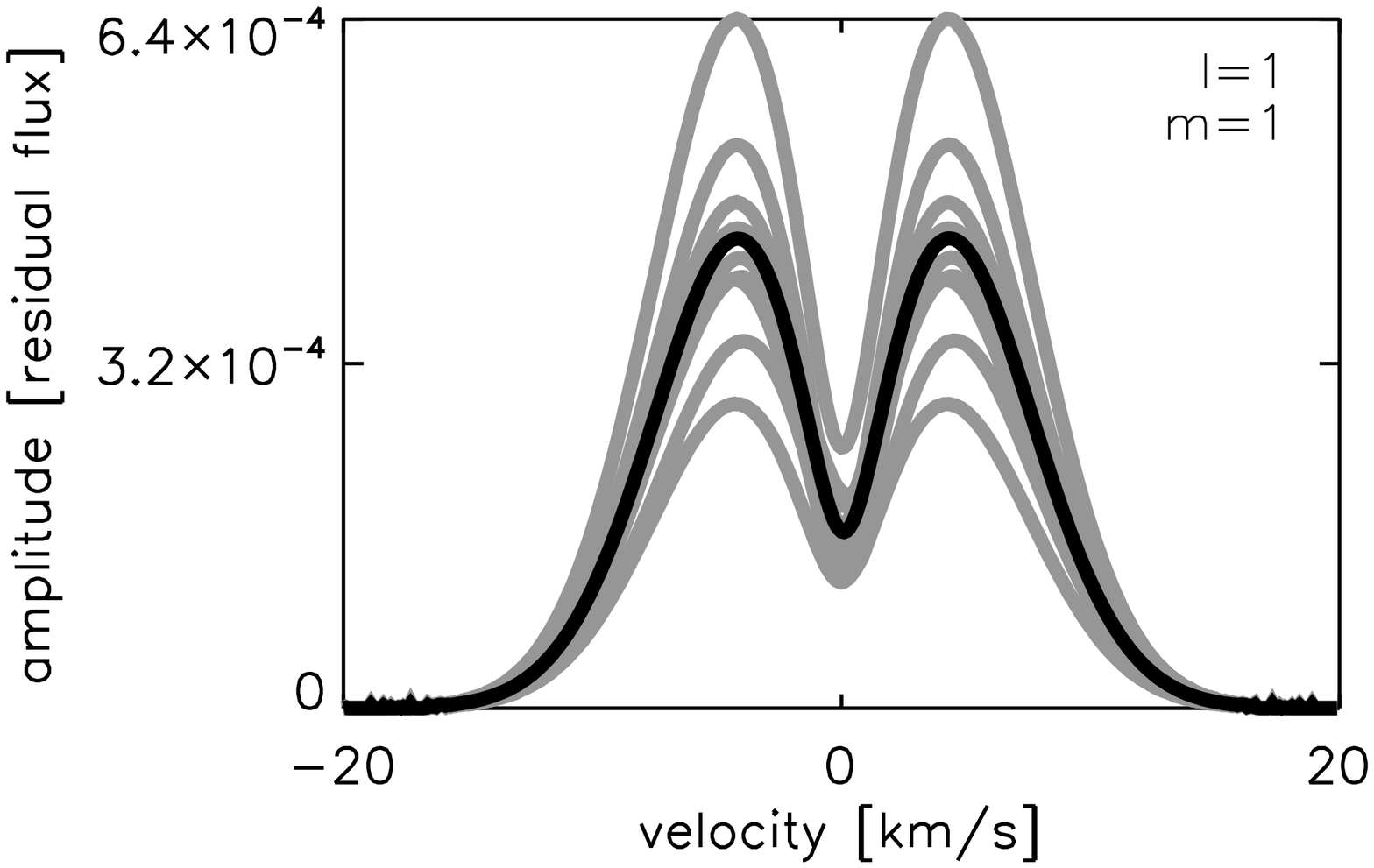}
\end{minipage}
\hfill
\begin{minipage}{4.2cm}
\centering
\includegraphics[width=4.2cm]{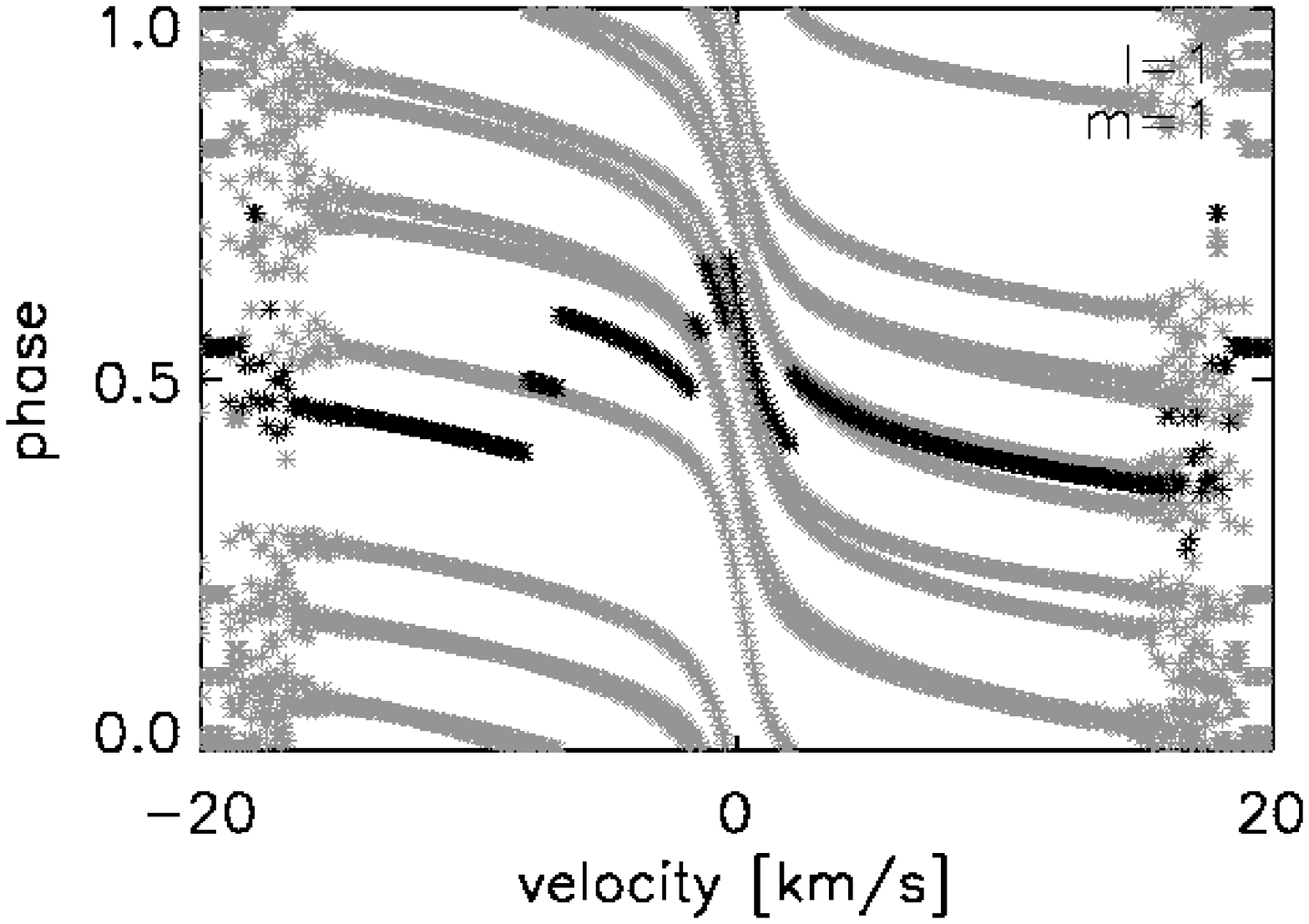}
\end{minipage}
\end{minipage}
\begin{minipage}{8.6cm}
\begin{minipage}{4.2cm}
\centering
\includegraphics[width=4.2cm]{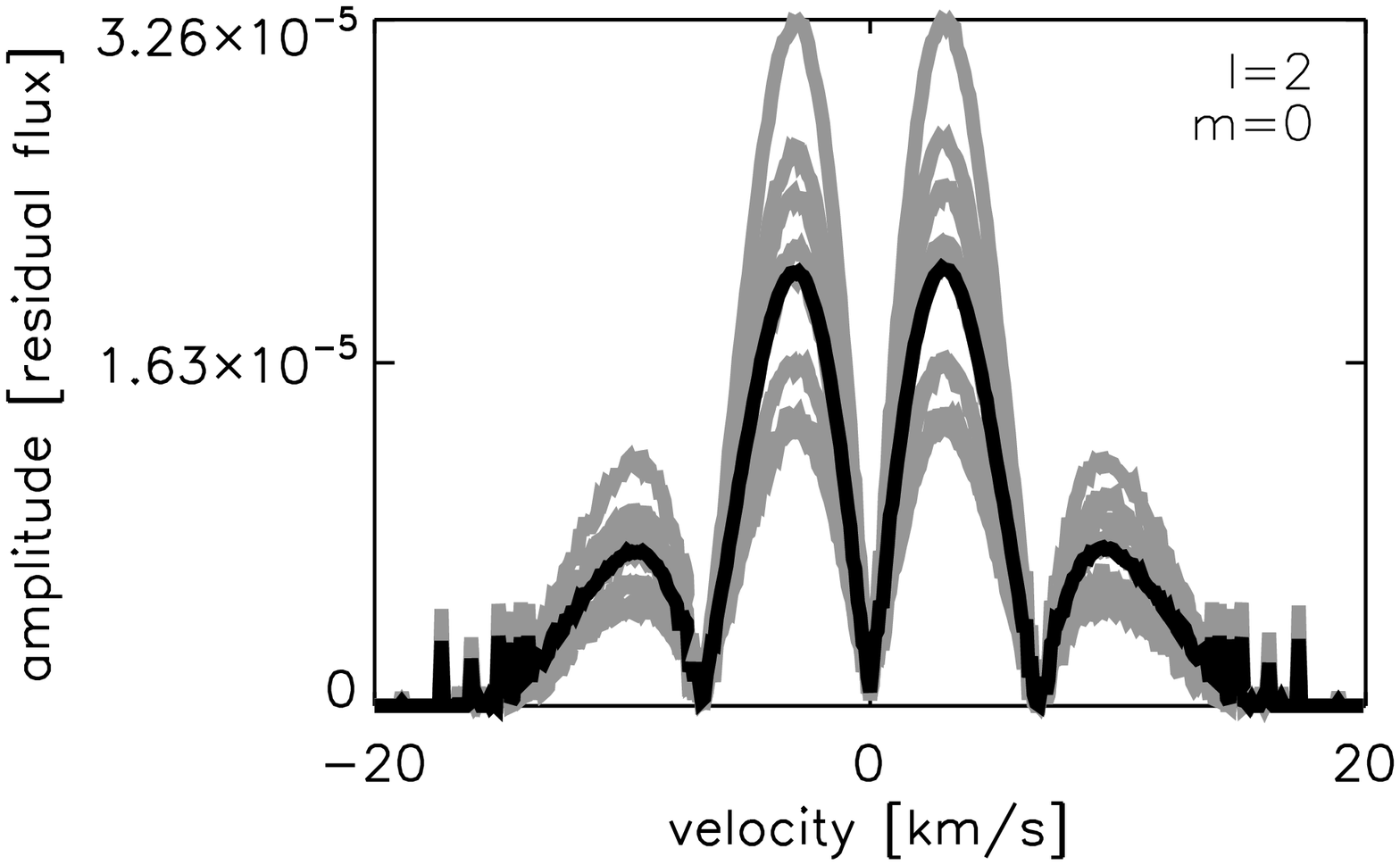}
\end{minipage}
\hfill
\begin{minipage}{4.2cm}
\centering
\includegraphics[width=4.2cm]{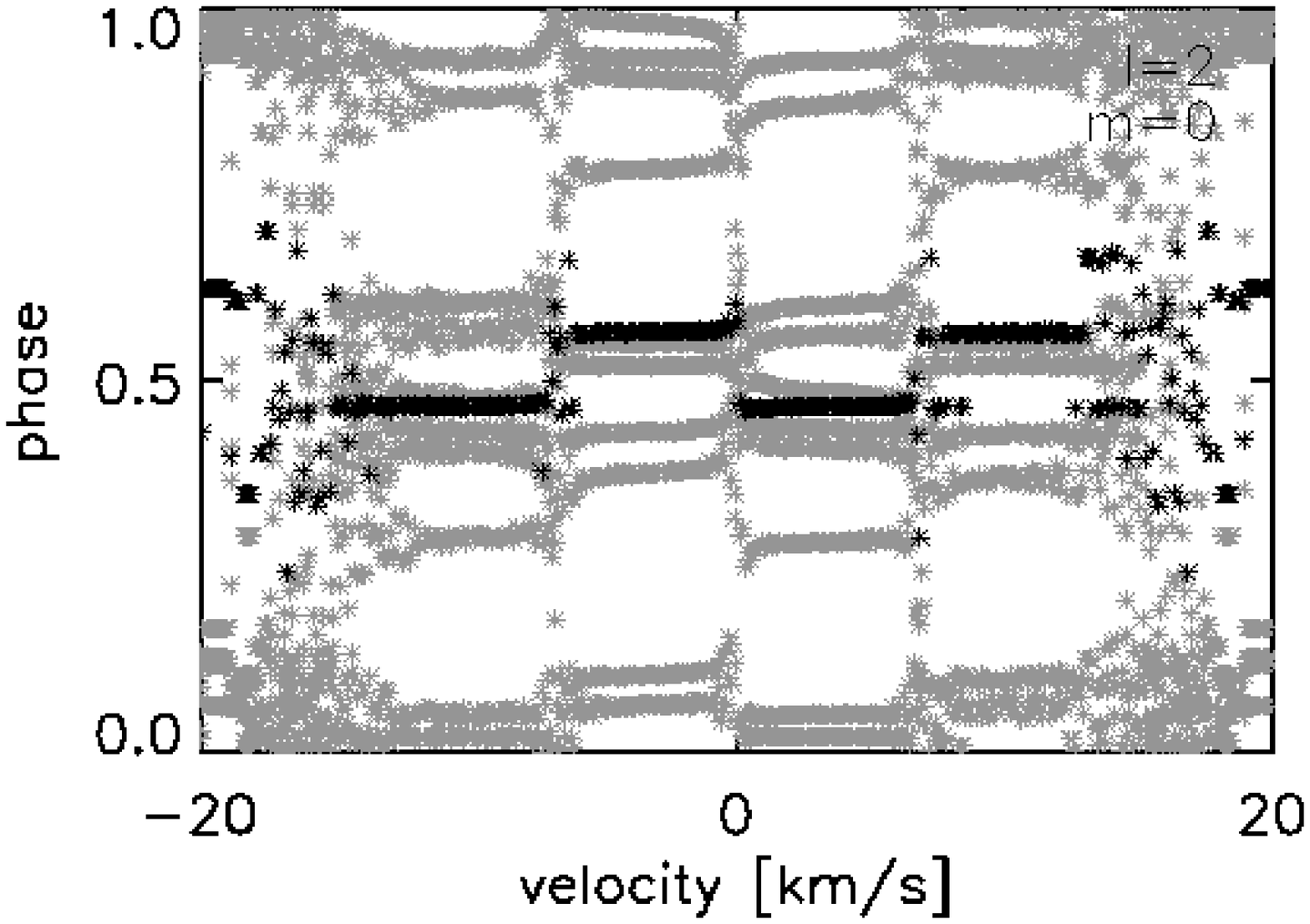}
\end{minipage}
\end{minipage}
\begin{minipage}{8.6cm}
\begin{minipage}{4.2cm}
\centering
\includegraphics[width=4.2cm]{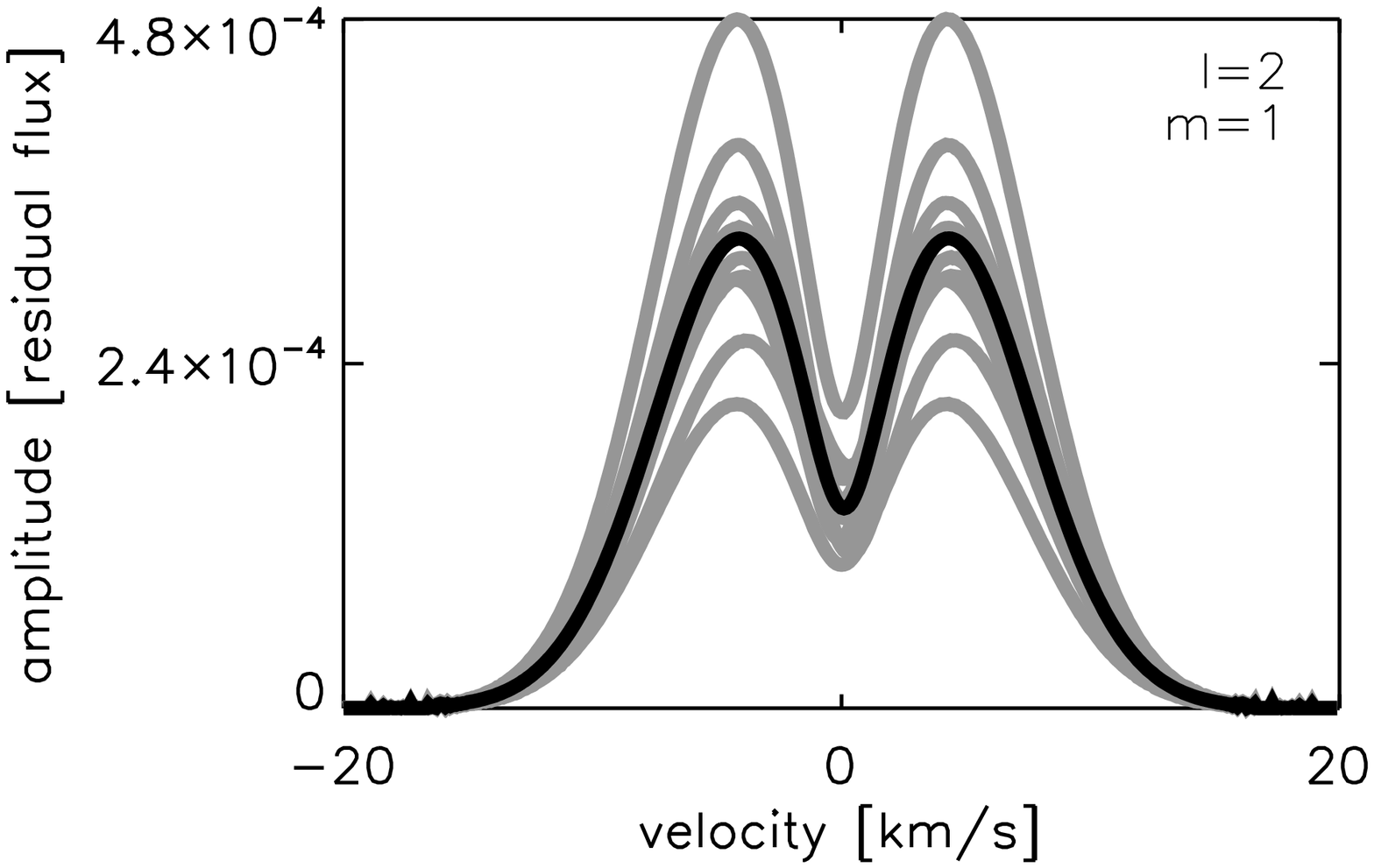}
\end{minipage}
\hfill
\begin{minipage}{4.2cm}
\centering
\includegraphics[width=4.2cm]{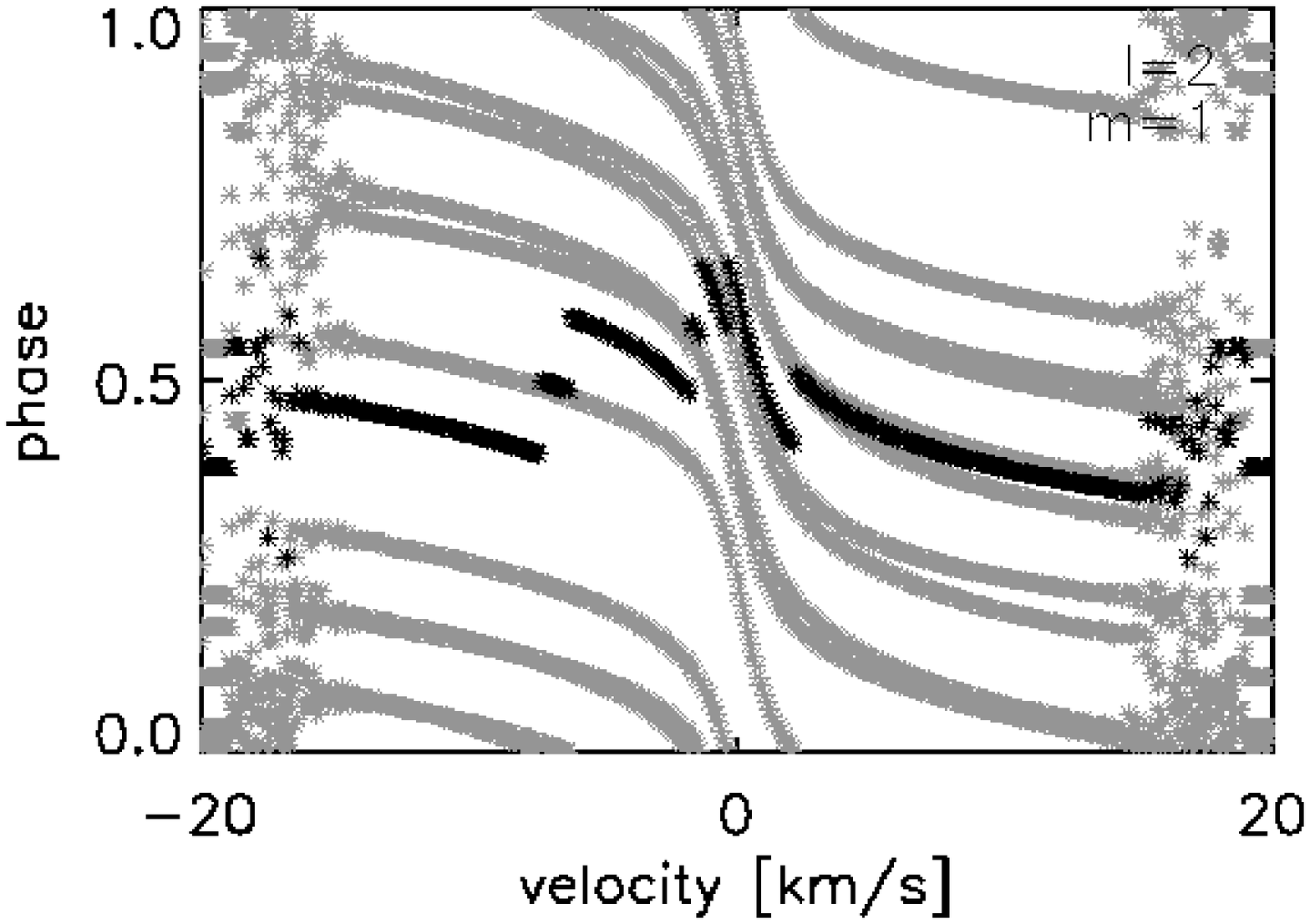}
\end{minipage}
\end{minipage}
\begin{minipage}{8.6cm}
\begin{minipage}{4.2cm}
\centering
\includegraphics[width=4.2cm]{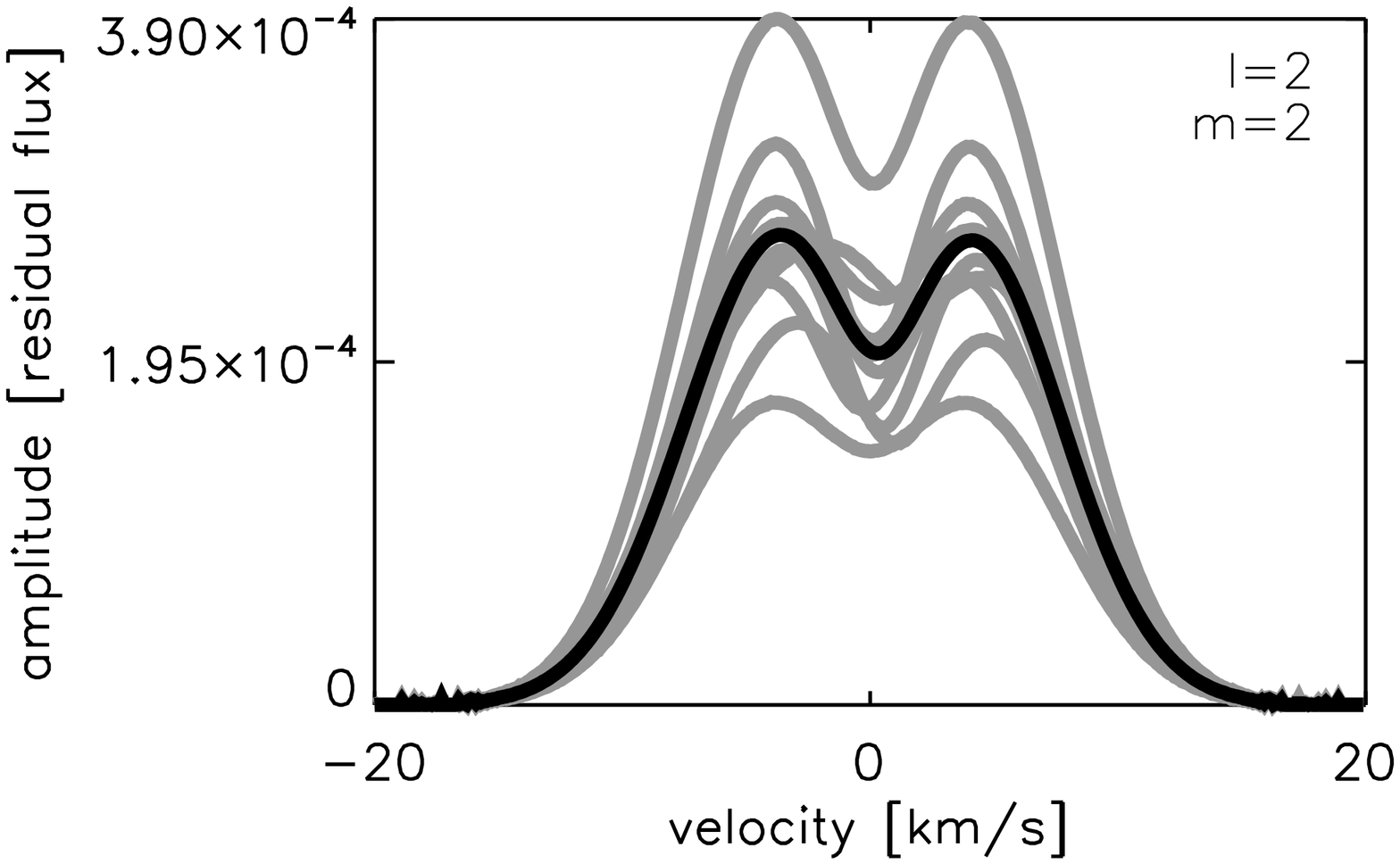}
\end{minipage}
\hfill
\begin{minipage}{4.2cm}
\centering
\includegraphics[width=4.2cm]{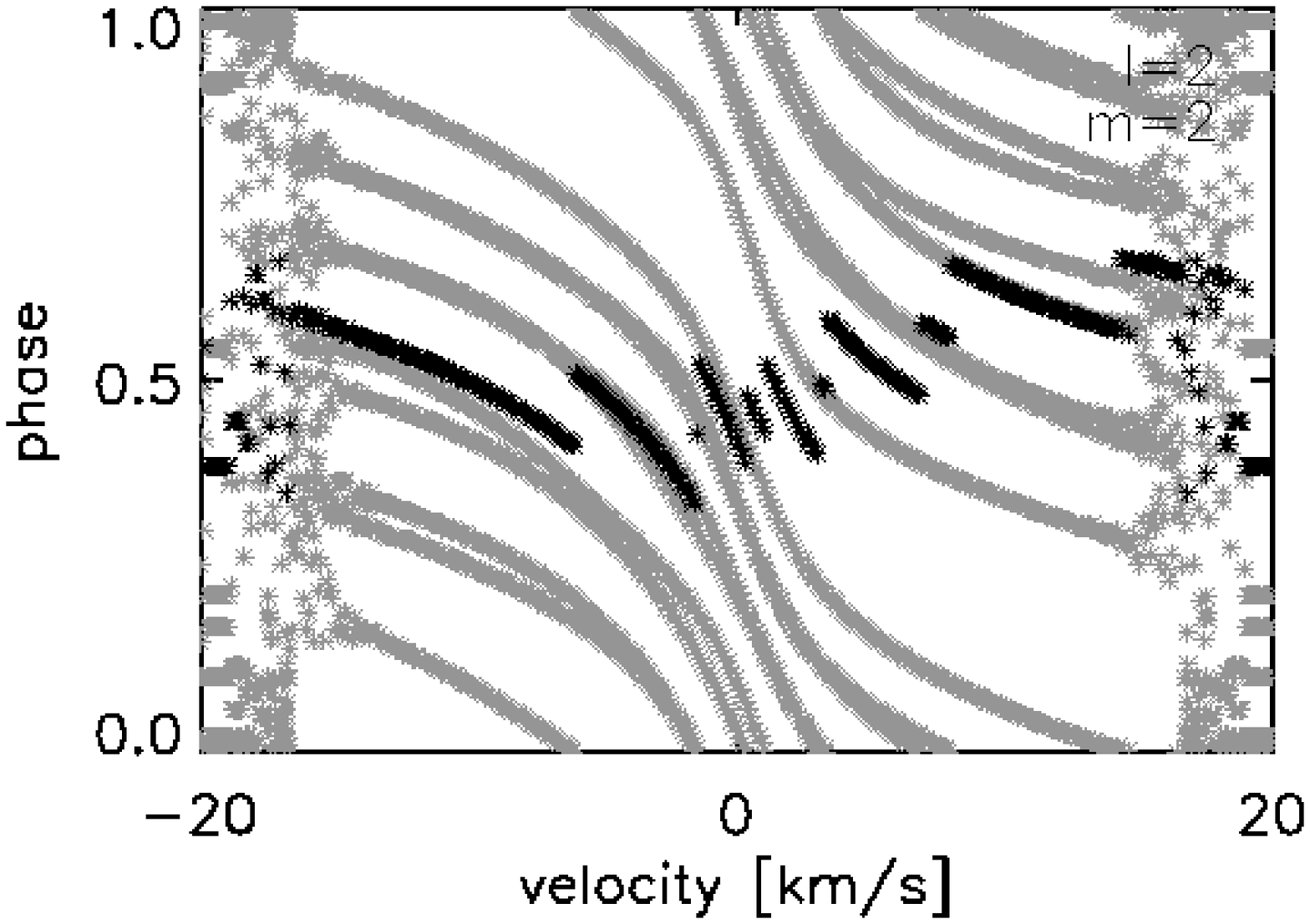}
\end{minipage}
\end{minipage}
\caption{Amplitude and phase distributions for simulated line profiles with a two day
damping time at an inclination angle of $i=55^{\circ}$. 
For each mode, ten different
realisations are shown in grey and the average distribution is 
shown in black. Top left:
$\ell=0$, $m=0$, top right: $\ell=1$, $m=0$, middle left: $\ell=1$, $m=1$, middle right: $\ell=2$, $m=0$, bottom left: $\ell=2$, $m=1$, bottom right: $\ell=2$,
$m=2$.}
\label{amplphaseveleta2i55}
\end{figure*}

\begin{figure*}
\begin{minipage}{8.6cm}
\begin{minipage}{4.25cm}
\centering
\includegraphics[width=4.25cm]{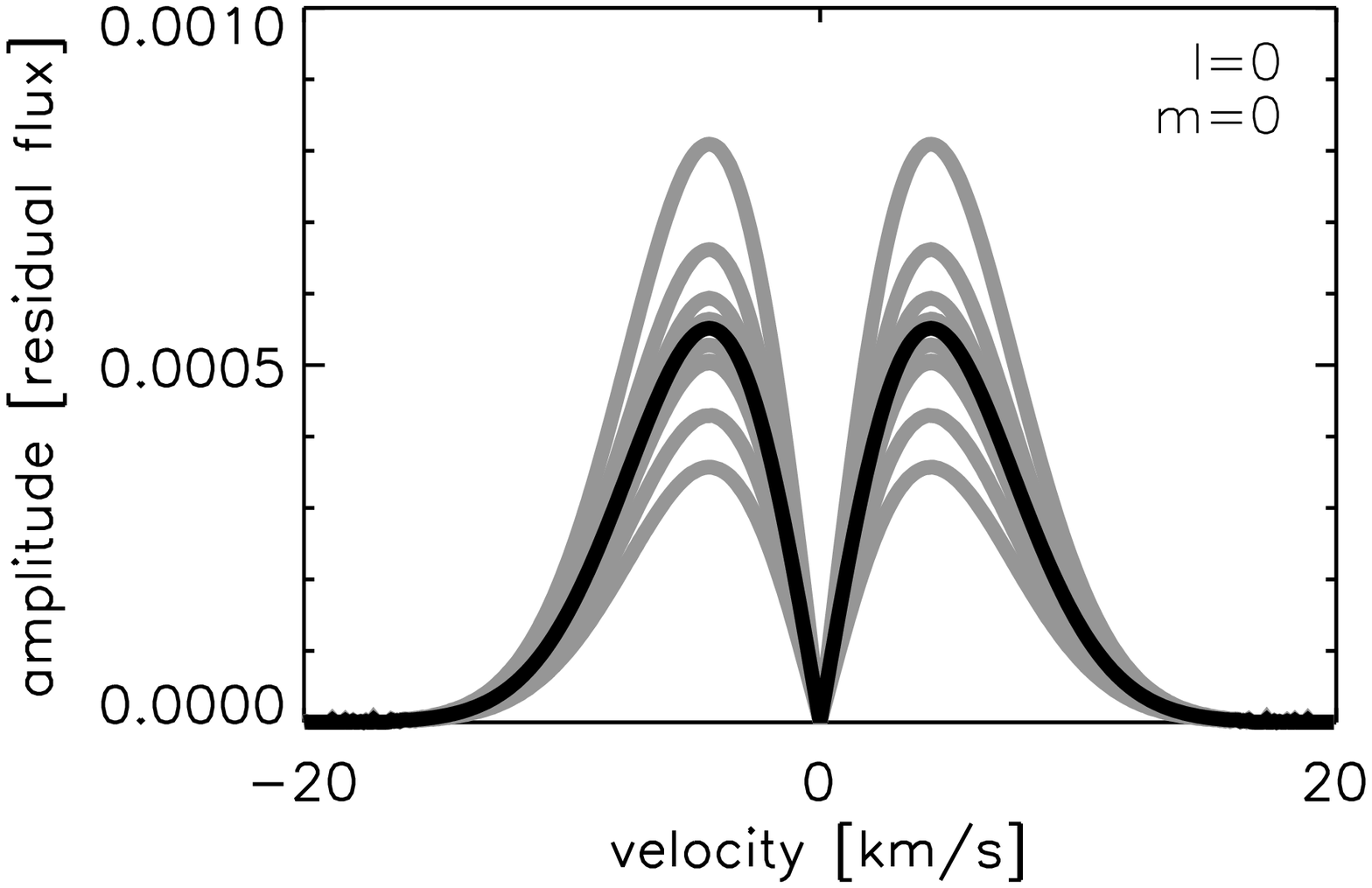}
\end{minipage}
\hfill
\begin{minipage}{4.25cm}
\centering
\includegraphics[width=4.25cm]{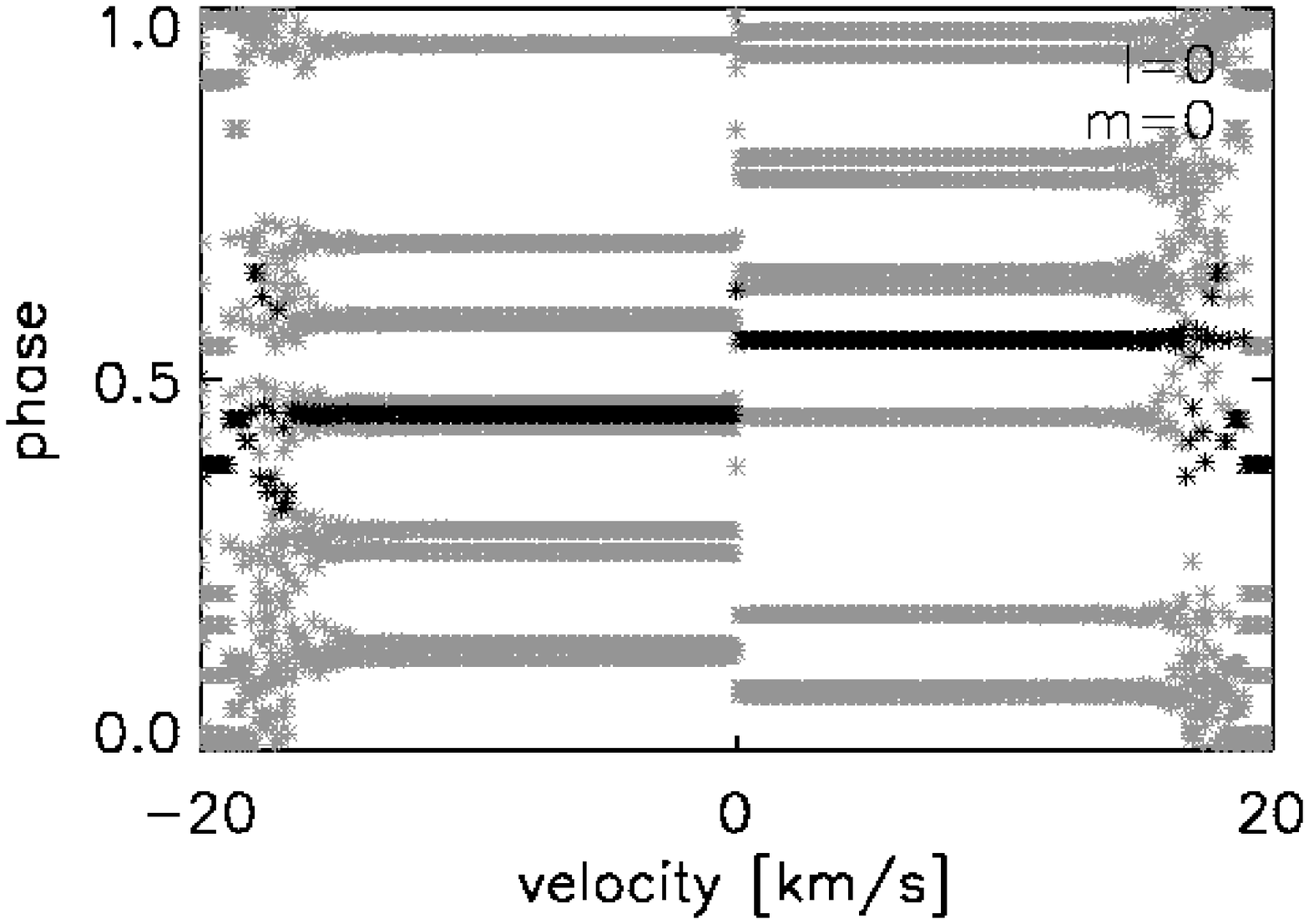}
\end{minipage}
\end{minipage}
\begin{minipage}{8.6cm}
\begin{minipage}{4.25cm}
\centering
\includegraphics[width=4.25cm]{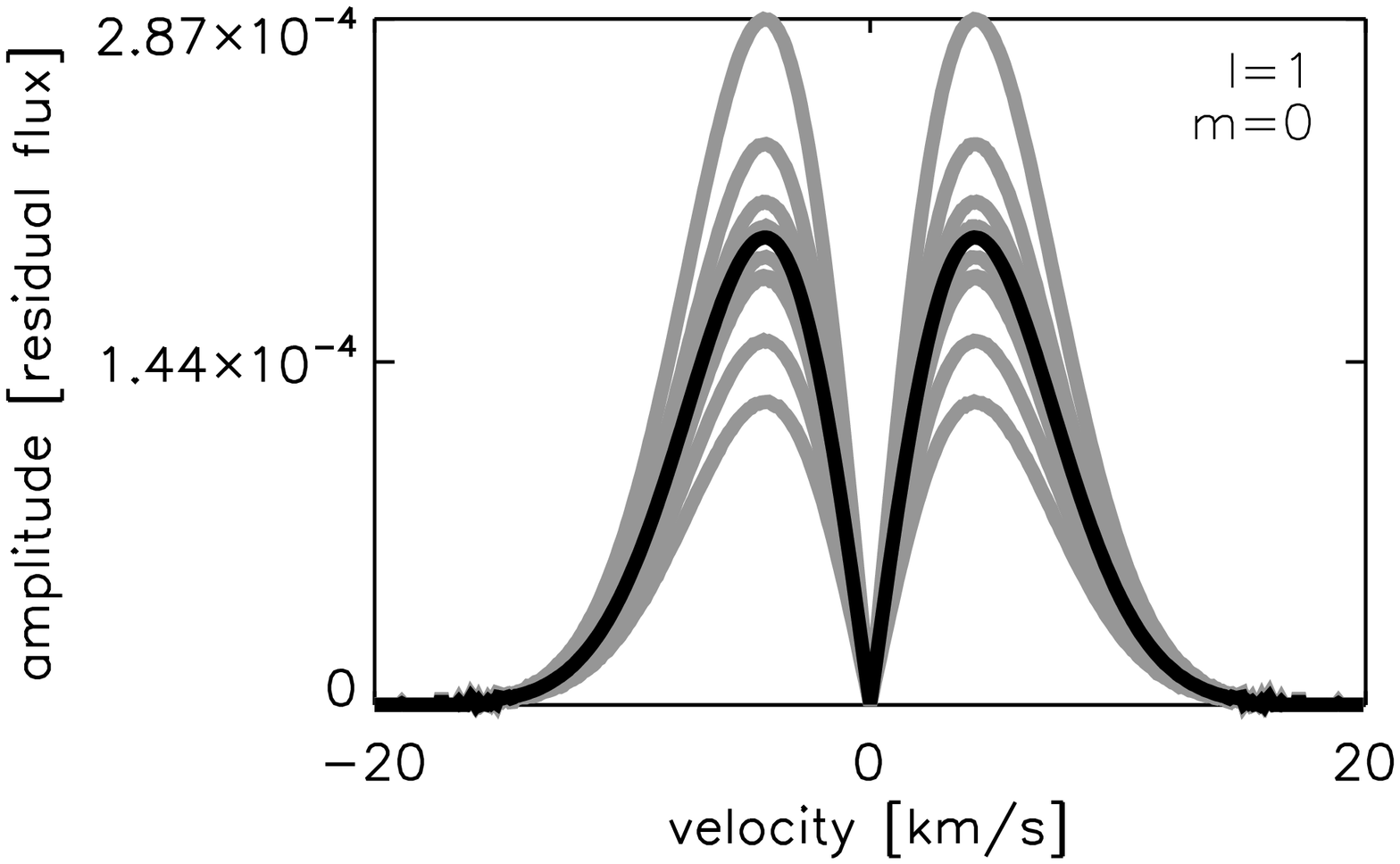}
\end{minipage}
\hfill
\begin{minipage}{4.25cm}
\centering
\includegraphics[width=4.25cm]{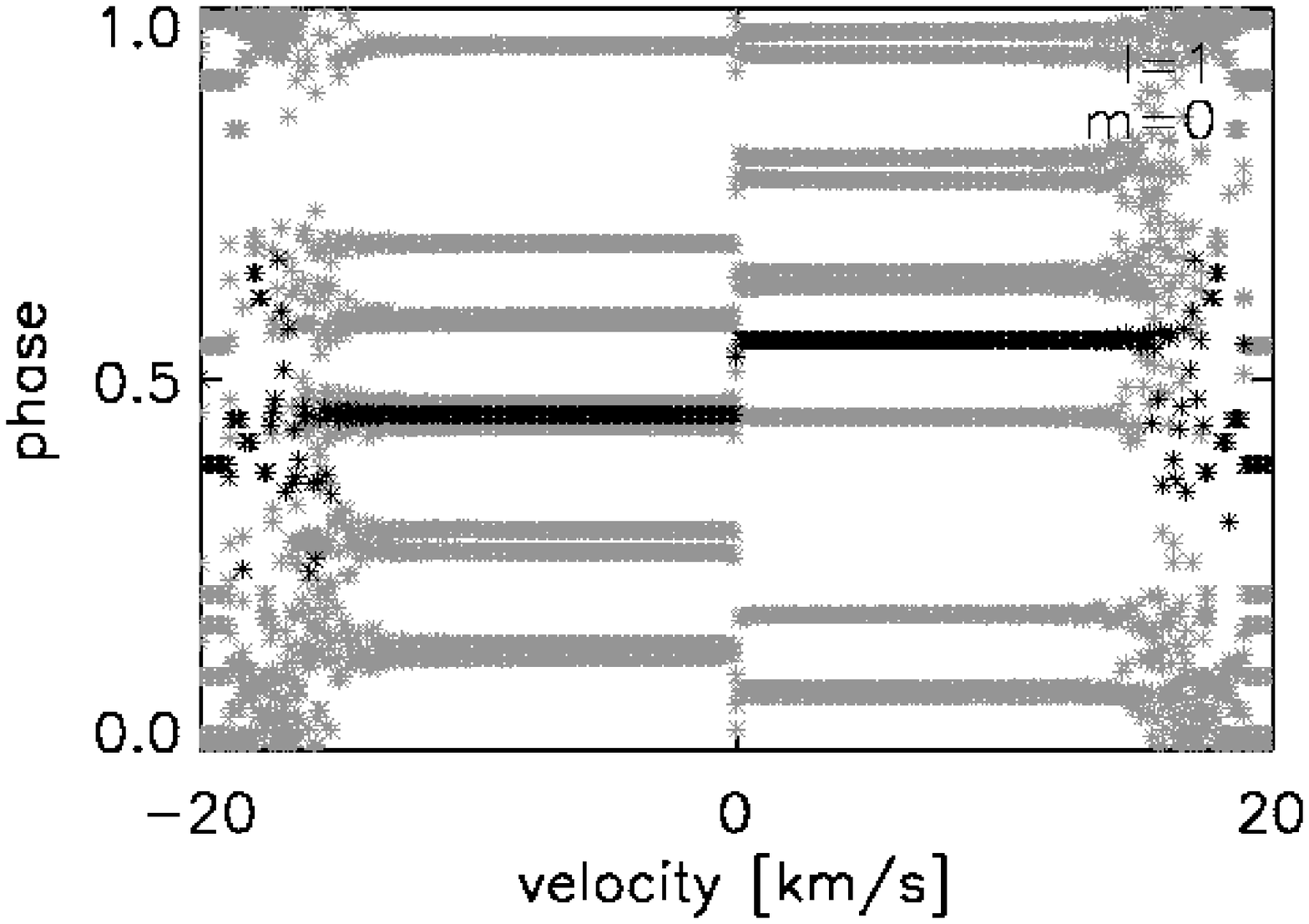}
\end{minipage}
\end{minipage}
\begin{minipage}{8.6cm}
\begin{minipage}{4.2cm}
\centering
\includegraphics[width=4.2cm]{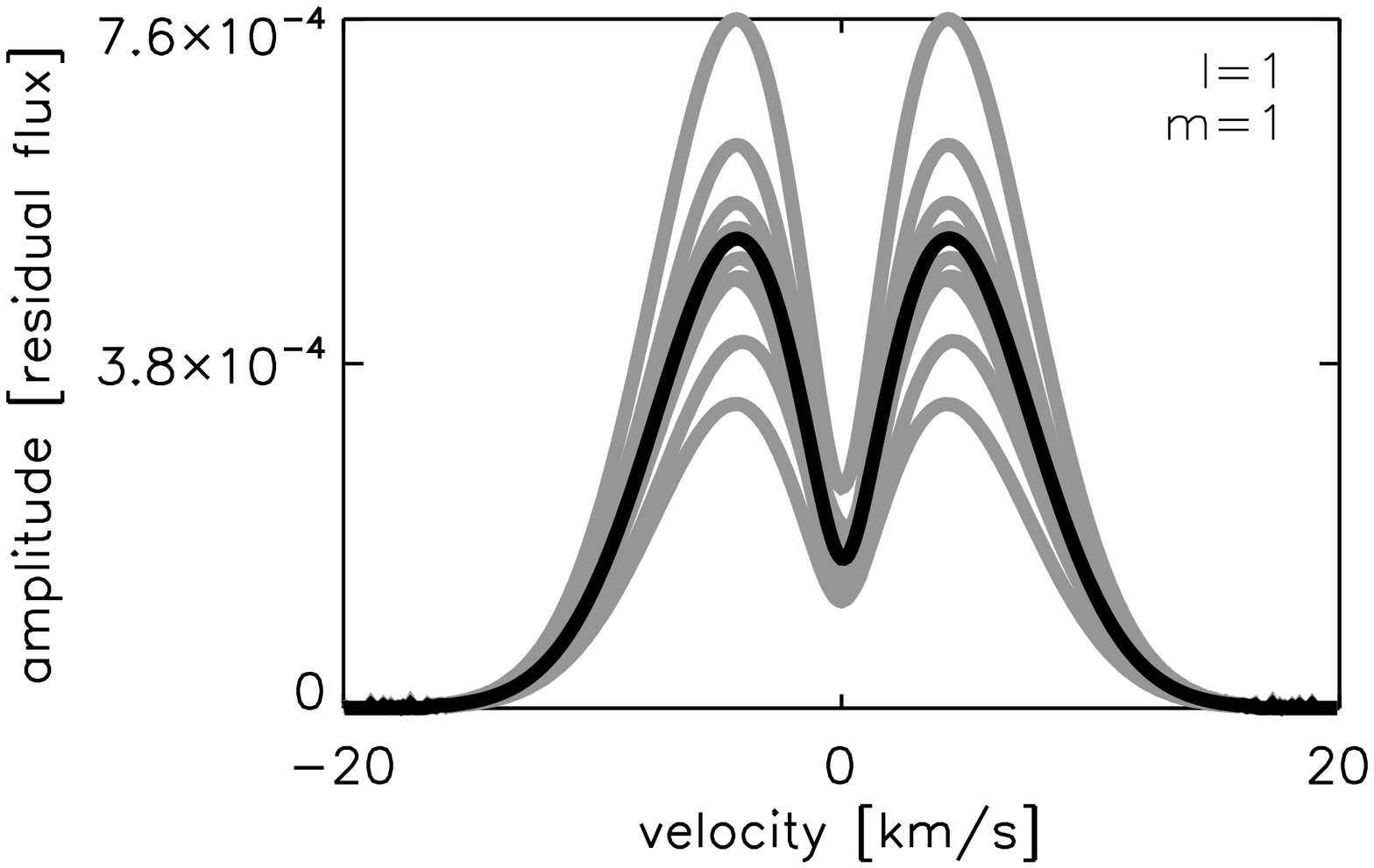}
\end{minipage}
\hfill
\begin{minipage}{4.2cm}
\centering
\includegraphics[width=4.2cm]{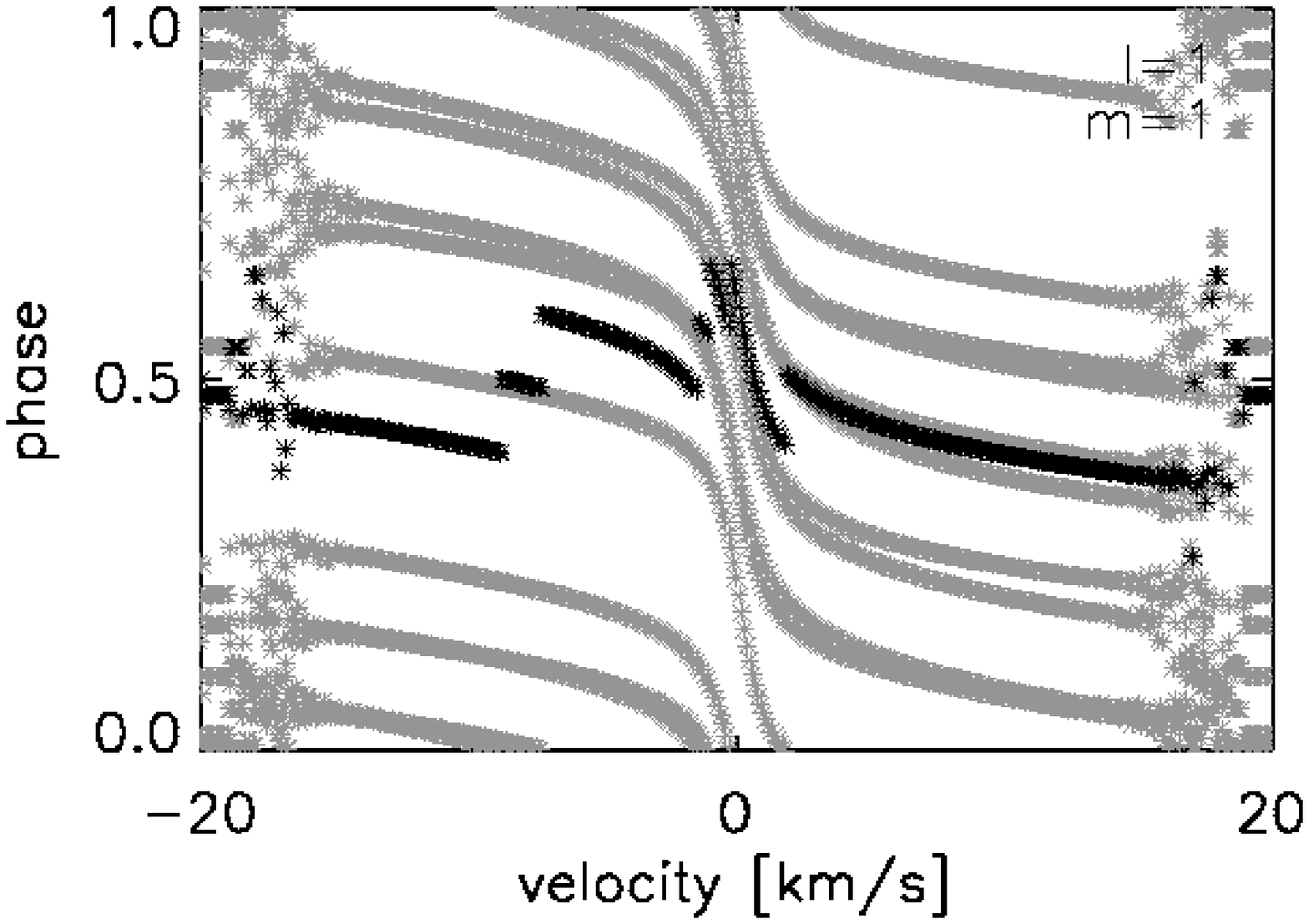}
\end{minipage}
\end{minipage}
\begin{minipage}{8.6cm}
\begin{minipage}{4.2cm}
\centering
\includegraphics[width=4.2cm]{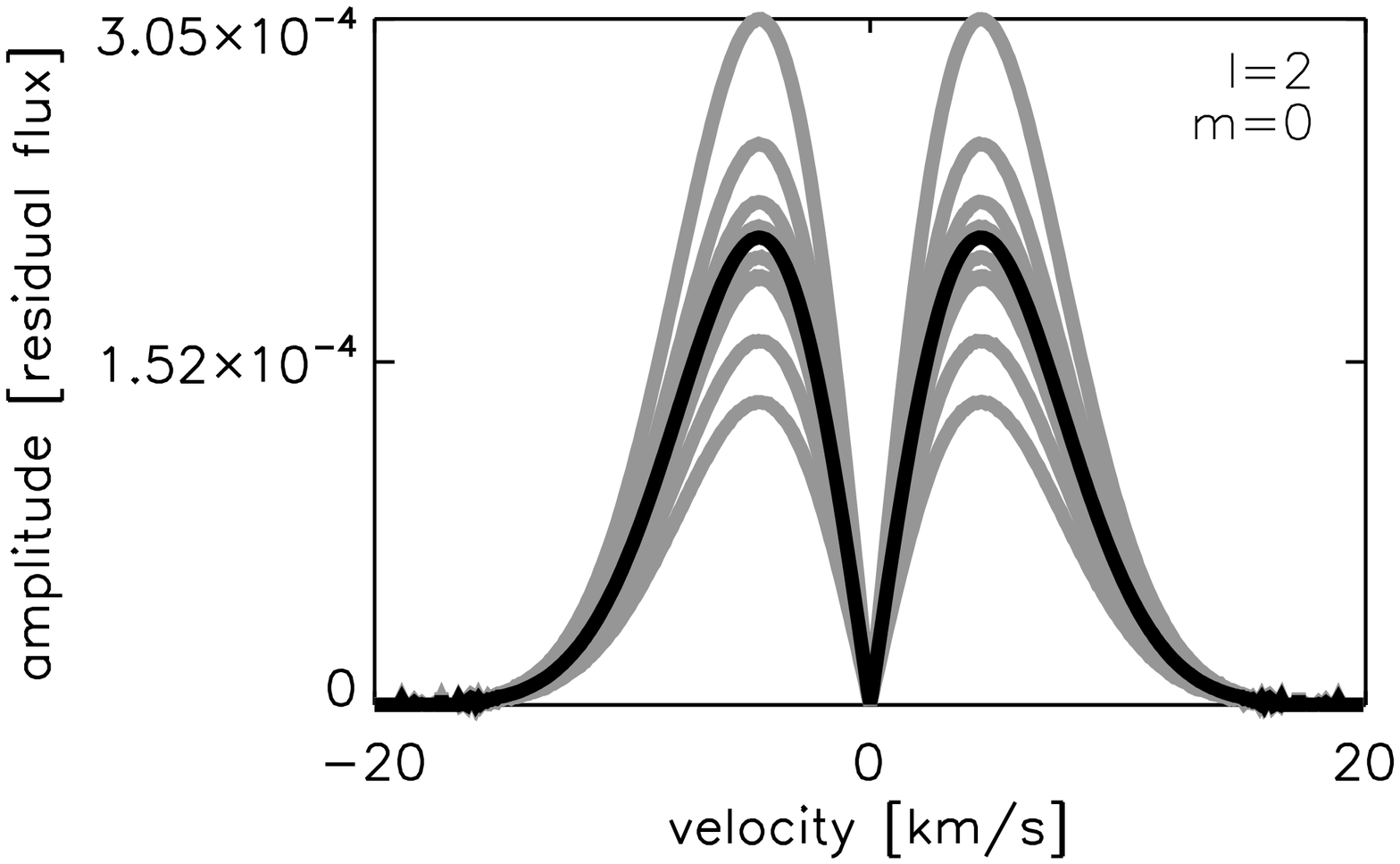}
\end{minipage}
\hfill
\begin{minipage}{4.2cm}
\centering
\includegraphics[width=4.2cm]{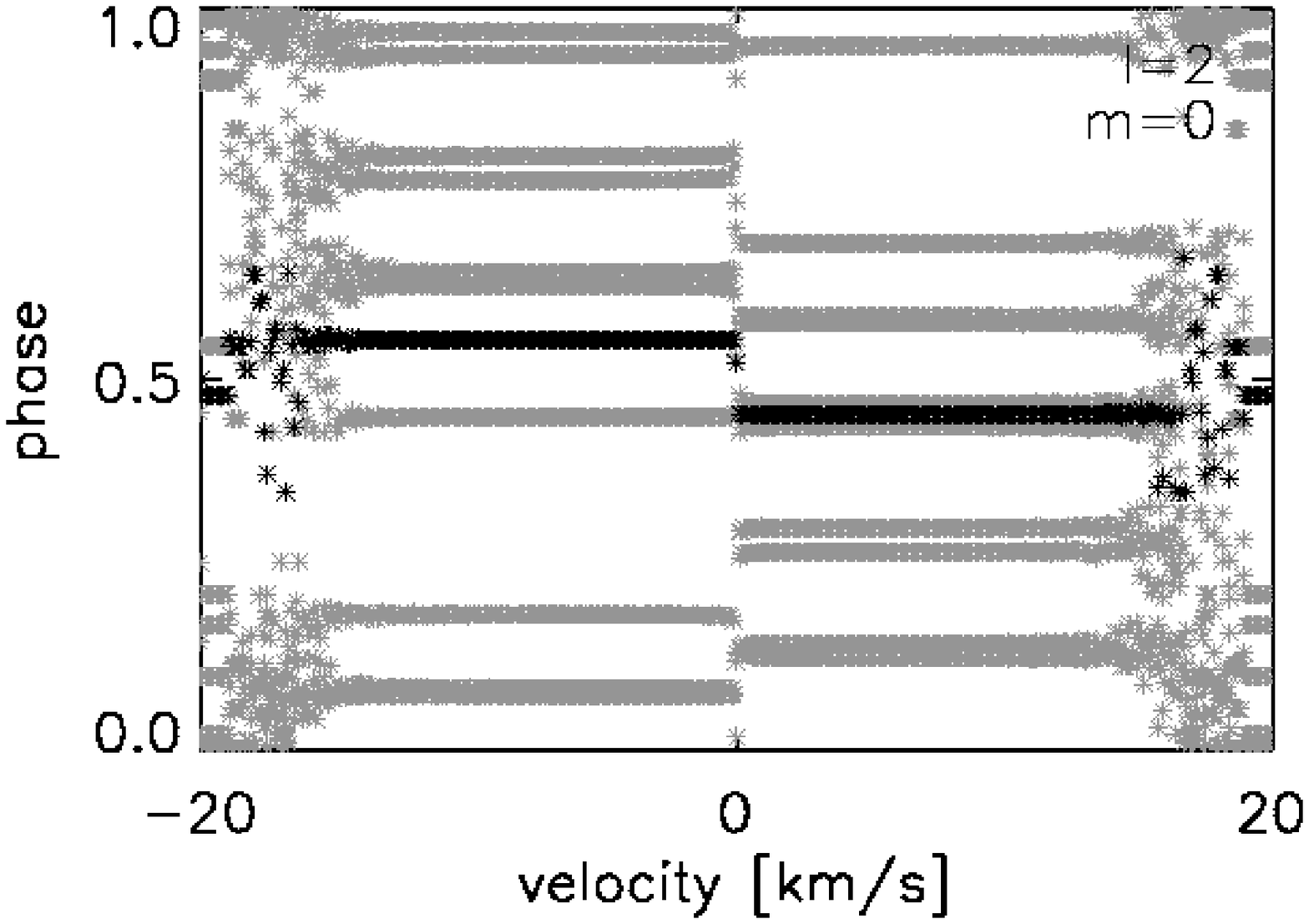}
\end{minipage}
\end{minipage}
\begin{minipage}{8.6cm}
\begin{minipage}{4.2cm}
\centering
\includegraphics[width=4.2cm]{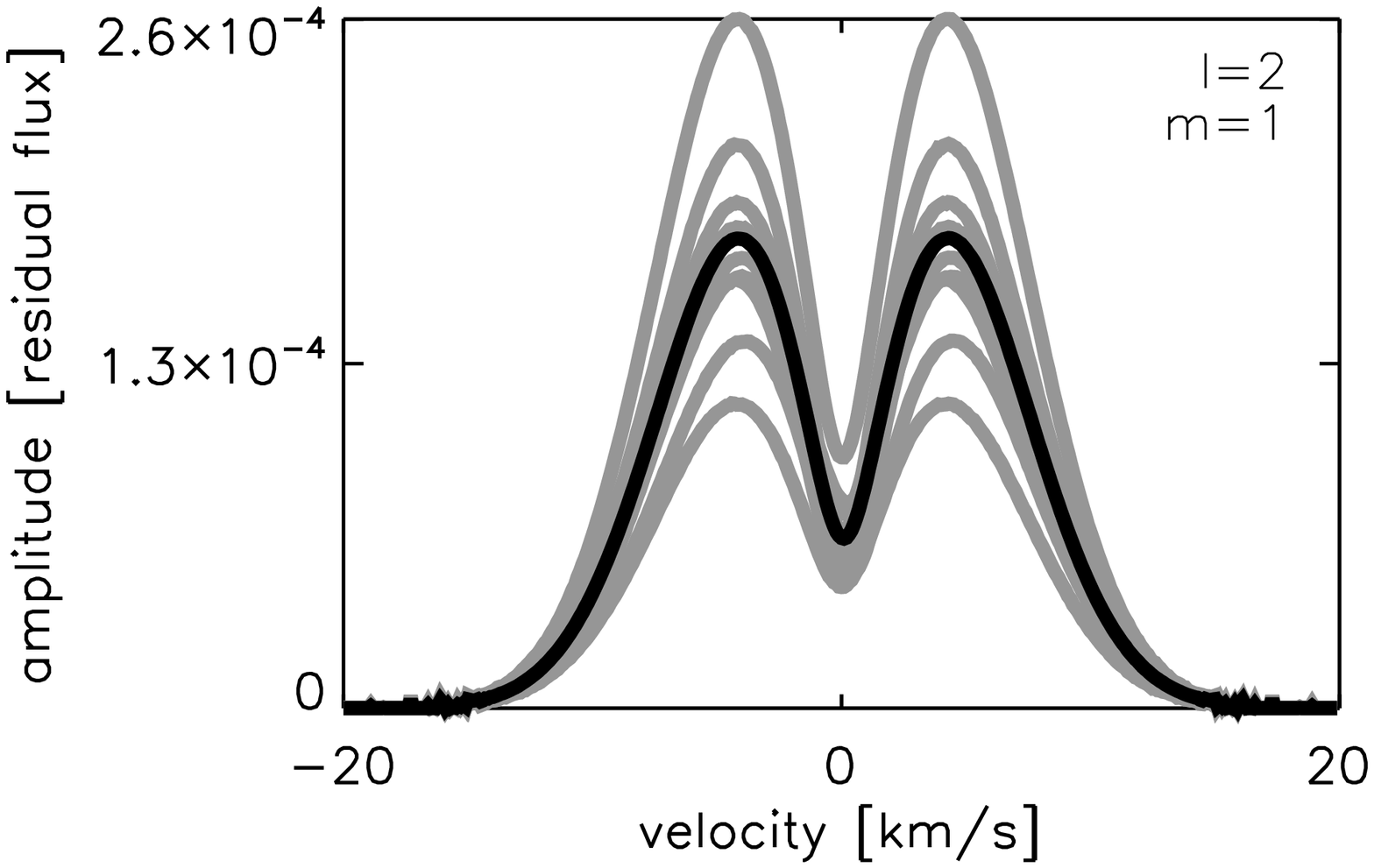}
\end{minipage}
\hfill
\begin{minipage}{4.2cm}
\centering
\includegraphics[width=4.2cm]{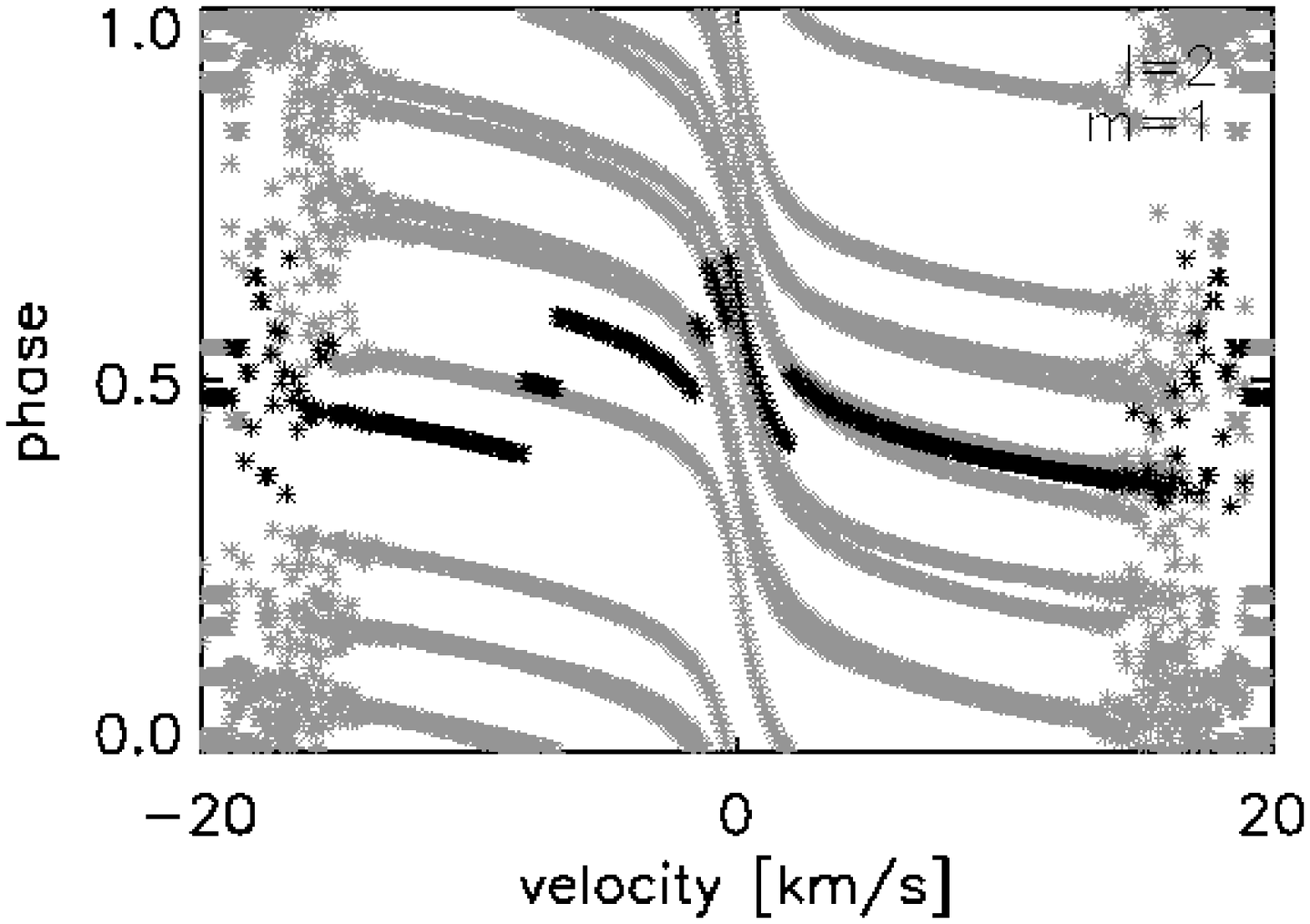}
\end{minipage}
\end{minipage}
\begin{minipage}{8.6cm}
\begin{minipage}{4.2cm}
\centering
\includegraphics[width=4.2cm]{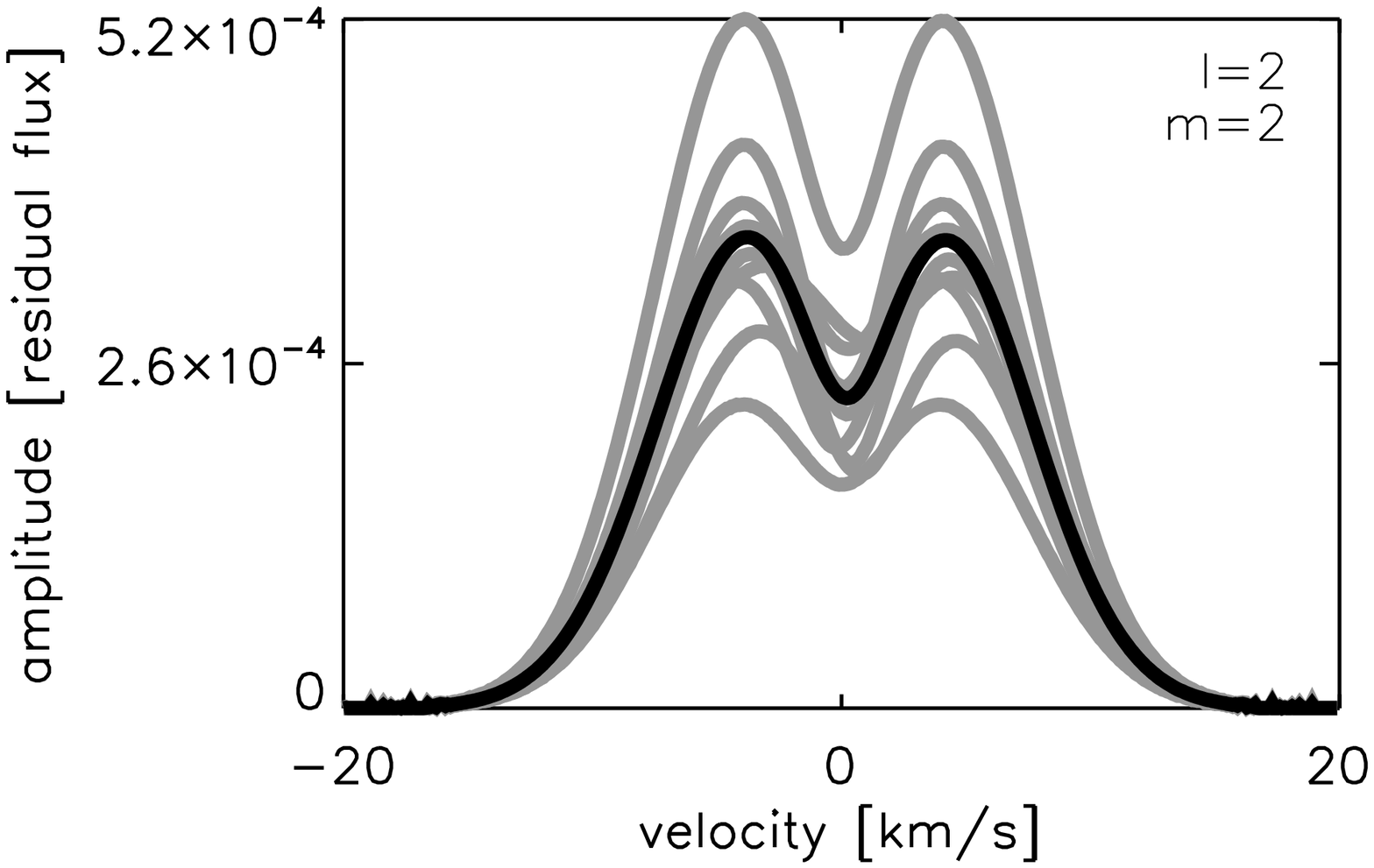}
\end{minipage}
\hfill
\begin{minipage}{4.2cm}
\centering
\includegraphics[width=4.2cm]{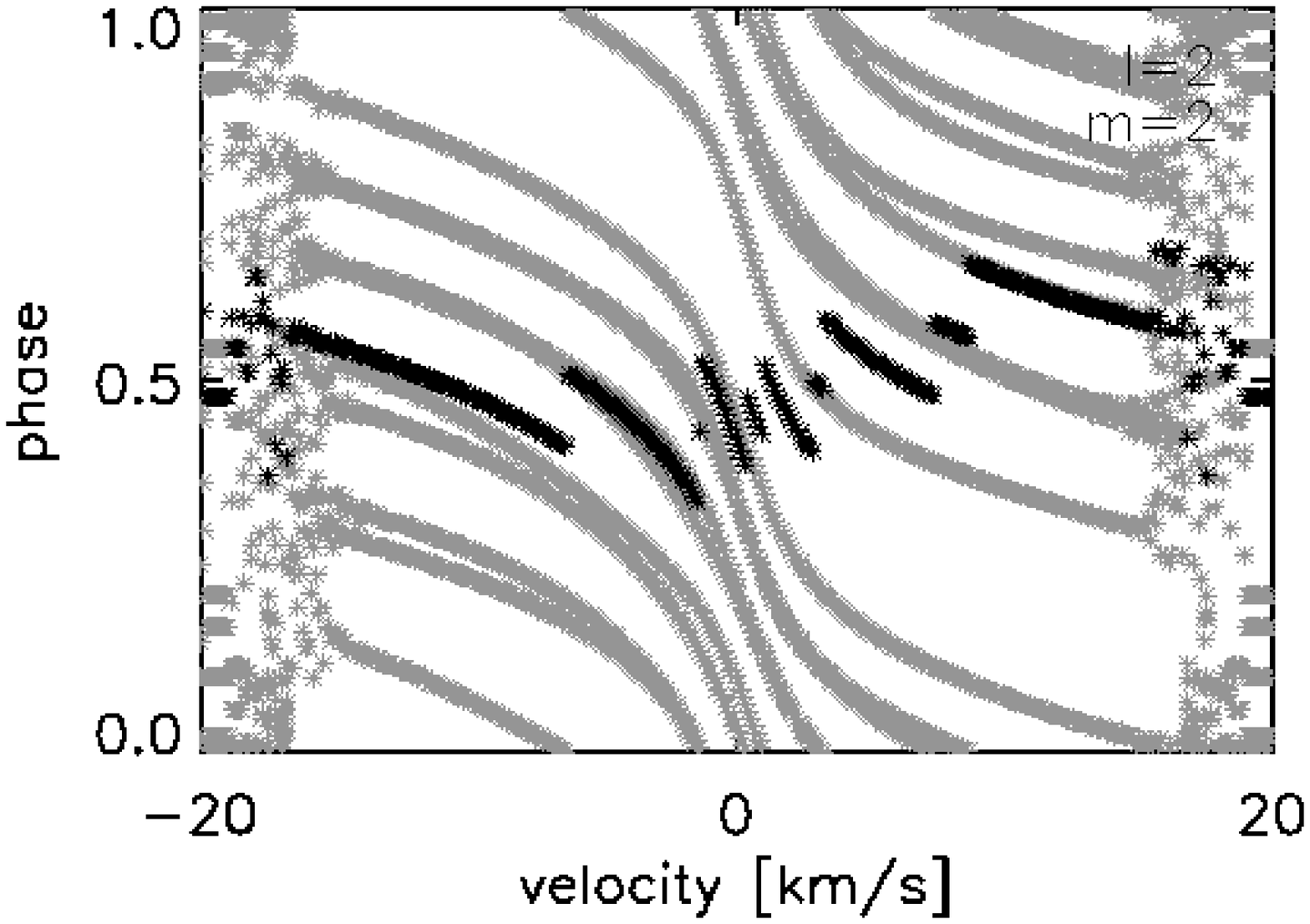}
\end{minipage}
\end{minipage}
\caption{Amplitude and phase distributions for simulated line profiles with a two day
damping time at an inclination angle of $i=75^{\circ}$. 
For each mode, ten different
realisations are shown in grey and the average distribution is 
shown in black. Top left:
$\ell=0$, $m=0$, top right: $\ell=1$, $m=0$, middle left: $\ell=1$, $m=1$, middle right: $\ell=2$, $m=0$, bottom left: $\ell=2$, $m=1$, bottom right: $\ell=2$,
$m=2$.}
\label{amplphaseveleta2i75}
\end{figure*}

The results obtained in the previous section are not anticipated from a
theoretical viewpoint (e.g. \cite{dziembowski2001}). One may therefore wonder
if the profiles simulated for infinite mode lifetimes differ too much from
reality, i.e.\ from those for finite lifetimes as in the data.  The damping and
re-excitation times for red giant oscillations turn out to be short. Indeed,
\cite{stello2004} derived an oscillation mode lifetime for $\xi$ Hydrae
(G7III) of only approximately two days. In order to take this into account, and
to investigate the robustness of our conclusion against finite lifetimes, we
computed line profile variations simulated for stochastically excited
modes. Such line profile variations are not yet available in the literature. We
provide them here in an attempt to make a definite conclusion on the nature of 
$\epsilon$ Ophiuchi's dominant oscillation modes.

A damped and re-excited oscillation mode is damped by a factor $e^{-\eta t}$,
with $\eta$ the damping rate, and re-excited before it is able to damp out. As a
consequence both the amplitude and the phase of the oscillation are time
dependent.
\begin{equation}
f(t)=A(t)\sin(2 \pi \nu t+ \psi (t)).
\label{sinos}
\end{equation} 
To simulate such an oscillator we follow the description of
\cite{deridder2006a}, and we compute
\begin{equation}
f(t)= B(t) \sin(2\pi\nu t)+ C(t) \cos(2\pi\nu t),
\label{dampreex}
\end{equation}
where we let the amplitudes $B$ and $C$ vary with a first order autoregressive
process in a discrete time domain with time step $\Delta t$:
\begin{equation}
B_{n}= e^{-\eta\Delta t} B_{n-1} + \varepsilon_{n+1},
\end{equation}
where $\varepsilon_{n+1}$ is a Gaussian distributed excitation kick. 
For more details we refer to \cite{deridder2006a}. The above is applied
to the equations for the pulsation velocity which are obtained by taking
the time derivative of the displacement components mentioned in Section 3.2,
Eq.(2) of \cite{deridder2002}. In case rotation is neglected, the three
spherical components of the pulsation velocity, i.e. $\upsilon_{r}$,
$\upsilon_{\theta}$ and $\upsilon_{\varphi}$, for the damped and re-excited case
become:
\begin{eqnarray}
\upsilon_{r}  =  -\upsilon_{p}N_{\ell}^{m}P_{\ell}^{m}(\cos\theta)e^{-\eta(t-n\Delta t_{kick})}*
\nonumber\\
 (B_{n}\sin(m\varphi+2 \pi \nu t) + C_{n}\cos(m\varphi +2 \pi\nu t)),
\label{vrad}
\end{eqnarray}
\begin{eqnarray}
\upsilon_{\theta}  = 
-K\upsilon_{p}N_{\ell}^{m}\frac{\partial}{\partial\theta}(P_{\ell}^{m}
(\cos\theta))e^{-\eta(t-n\Delta t_{kick})}*
\nonumber\\
(B_{n}\sin(m\varphi+2 \pi \nu t) +C_{n}\cos(m\varphi +2 \pi\nu t)),
\label{vtheta}
\end{eqnarray}
\begin{eqnarray}
\upsilon_{\varphi}  = 
-mK\upsilon_{p}N_{\ell}^{m}\frac{1}{\sin\theta}P_{\ell}^{m}(\cos\theta)e^{-\eta(t-n\Delta t_{kick})}*
\nonumber\\
 (B_{n}\cos(m\varphi+2 \pi \nu t) -C_{n}\sin(m\varphi +2 \pi\nu t)),
\label{vfi}
\end{eqnarray} 
with $\upsilon_{p}$ proportional to the pulsation amplitude, $N_{\ell}^{m}$ the
normalisation factor for the spherical harmonics $Y_{\ell}^{m}(\theta,\varphi)
\equiv P_{\ell}^{m}(\cos\theta)e^{im\varphi}$ and $K$ the ratio of the
horizontal to the vertical velocity amplitude.

Line profiles of oscillations with different wavenumbers ($\ell,m$), including
damping and re-excitation, are simulated at the observation times of $\epsilon$
Ophiuchi and according to its observed amplitudes. Subsequently, $\langle
\mathrm{v} \rangle$ is determined for these profiles in the same way as for the
observations.  The amplitude and phase distributions are then computed for the dominant
oscillation frequencies, as described in section 4.3. 

Only in case of an infinite number of observations over an infinite timespan,
the real dominant frequency can be obtained for a damped and re-excited
oscillator. Smaller samples of observations, single realisations, are all
different and have their own dominant frequency and amplitude distribution. In
Figures~\ref{amplphaseveleta2i55} and \ref{amplphaseveleta2i75}, simulated amplitude and phase distributions for different
wavenumbers are plotted at inclination angles $i=55^{\circ}$ and $i=75^{\circ}$, respectively. For each mode, ten single realisations are plotted in
grey, with the average plotted in black. From these Figures it becomes clear that,
although all realisations are different from each other, it is still possible from the amplitude distributions, to
distinguish between different $m$ values, just by visual inspection. In
particular, the different realisations for $m=0$ modes all lead to zero
amplitude in the line centre. In contrast with the pronounced shapes of the phase diagrams of oscillations with infinite lifetimes (see e.g. right bottom panel Figures~\ref{ampphasescheml0m0} and \ref{ampphasescheml2m2} or \cite{hekker2006}), Figures~\ref{amplphaseveleta2i55} and \ref{amplphaseveleta2i75} show that the phase distributions across a profile can be quite different for each realisation due to damping and re-excitation. Therefore, phase distributions across a profile of oscillations with finite mode lifetimes can not yet be explored in terms of mode identification. These results are robust against a change in
inclination angle, as can be seen by comparing Figures~\ref{amplphaseveleta2i55} and \ref{amplphaseveleta2i75}
with a similar plot for a different inclination provided in \cite{hekker2006}.

\section{Conclusions}

We have found a different amplitude distribution across the cross correlation
profile for the dominant frequencies of the pulsating red giant star $\epsilon$
Ophiuchi.  This clearly indicates that modes with different wavenumbers
($\ell,m$) must be present in the data. Moreover, the amplitude distribution of
the dominant mode is clearly different from the one due to $m=0$ modes.  Thus,
we have detected non-radial modes in the observed cross correlation profiles of
this star.

We compared the observed amplitude distributions of $\epsilon$ Ophiuchi with the
one of simulated cross correlation profiles in an attempt to identify the
wavenumbers of the individual modes.  Due to the fact that single realisations
of different modes are in some cases comparable, e.g.\ for $(\ell,m)=(1,1)$ and
$(2,1)$ (see Figures\,\ref{amplphaseveleta2i55} and \ref{amplphaseveleta2i75}), different sets of wavenumbers are
equally likely for observed modes.  It is therefore clearly not yet possible to
identify the individual modes unambiguously with the method used in the present
work, but clear evidence for modes with $m \neq 0$ values is found.

\cite{hekker2006} also analysed the line profile variations of three additional
pulsating giants and came to similar conclusion than for $\epsilon$
Ophiuchi for two of these three stars. Moreover, non-radial mixed oscillation
modes were recently also found in a red giant observed with the MOST space
mission (\cite{matthews2006}, these proceedings). It therefore seems that red giant
stars pulsate in both radial and non-radial modes. These observational results
constitute a fruitful starting point to fine-tune theoretical predictions of
stochastically excited modes in evolved stars.

\section*{Acknowledgments}
SH wants to thank the staff at the Instituut voor Sterrenkunde at the Katholieke
Universiteit Leuven for their hospitality during her three month visit. JDR is a
postdoctoral fellow of the fund for Scientific Research, Flanders. FC was
supported financially by the Swiss National Science Foundation. The authors are
supported by the Fund for Scientific Research of Flanders (FWO) under grant
G.0332.06 and by the Research Council of the University of Leuven under grant
GOA/2003/04.

\bibliography{bibSOHO}

\end{document}